\documentclass[journal]{IEEEtran}

\usepackage{ifpdf}
\usepackage{cite}
\ifCLASSINFOpdf
\else
\fi
\usepackage{amsmath}
\usepackage{algorithmic}
\usepackage{xcolor}
\usepackage{adjustbox}
\usepackage{array}
\usepackage[version=4]{mhchem}
\usepackage{booktabs}

 \usepackage[caption=false,font=footnotesize]{subfig}
\usepackage{comment}  % 加载 comment 宏包

\usepackage{fixltx2e}
\usepackage{stfloats}
\usepackage{bm}
\usepackage{graphicx}
\graphicspath{{images/}}
\usepackage{cite}
\usepackage{picinpar}
\usepackage{amsmath}
\usepackage{pifont}
\usepackage{enumerate}
\usepackage{siunitx}
\usepackage{breakurl}
\usepackage{epstopdf}
\usepackage{pbox}
\usepackage{amsthm}
\usepackage{geometry}
\usepackage{lipsum}  
\usepackage{booktabs}
\usepackage{array,tabularx,booktabs,makecell}
\newcolumntype{C}{>{\centering\arraybackslash}X}
\newcommand{\cmark}{\ding{51}} % √
\newcommand{\xmark}{\ding{55}} % ×

\begin{document}
\title{Transient Power Allocation Control Scheme for Hybrid Hydrogen Electrolyzer–Supercapacitor System with Autonomous Inertia Response}
\author{Pengfeng Lin,~\IEEEmembership{Member,~IEEE,}
	Guangjie Gao, ~\IEEEmembership{Member,~IEEE,}
        Jianjun Ma, ~\IEEEmembership{Member,~IEEE,}
	Miao Zhu, ~\IEEEmembership{Senior Member,~IEEE,}
        Xinan Zhang,~\IEEEmembership{Senior Member,~IEEE,}
	Ahmed Abu-Siada,~\IEEEmembership{Senior Member,~IEEE,}
	% <-this % stops a space
	% \thanks{M. Shell was with the Department of Electrical and Computer Engineering, Georgia Institute of Technology, Atlanta, GA, 30332 USA e-mail: (see http://www.michaelshell.org/contact.html).}% <-this % stops a space
	\thanks{
		Pengfeng Lin, Guangjie Gao, Jianjun Ma, and Miao Zhu are with Shanghai Jiao Tong University, China (Email: linpengfeng@ieee.org, \{guangjie\_gao, j.j.ma, miaozhu\} @sjtu.edu.cn).
		
		Xinan Zhang is with The University of Western Australia, Australia (Email:	xinan.zhang@uwa.edu.au).
		
		Ahmed Abu-Siada is with Curtin University, Australia (Email: a.abusiada@curtin.edu.au).
		
		.}% <-this % stops a space
	%\thanks{Corresponding Author: Yu Wang, wang\_yu@ntu.edu.sg.}
}

\maketitle
{\color{black}
\begin{abstract}
This paper proposes a hybrid hydrogen electrolyzer–supercapacitor system (HHESS) with a novel control strategy for renewable-dominant power grids. The HHESS consists of alkaline electrolyzers (AEL), proton exchange membrane electrolyzers (PEMEL), and supercapacitor (SC). The interfacing inverter between HHESS and power grid is regulated by an inertia emulation control strategy. Upon HHESS, AEL is with conventional V-P droop, whereas PEMEL and SC are designated with the proposed dynamic integral droop and capacitive integral droop, respectively. Benefitting from the coordination of three droops, within the HHESS, high-frequency transient power components are autonomously handled by SC, middle-frequency power components are regulated by PEMEL, and low-frequency steady-state power is addressed by AEL, characterized by low operational costs and longer lifespan. SC delivers immediate transient power, significantly alleviating dynamic stress on electrolyzers and achieving autonomous state-of-charge recovery without requiring additional communication. Implementing SOC recovery control enables the SC to withstand up to ten times more charge–discharge cycles compared to an SC without SOC recovery. Furthermore, a large-signal mathematical model based on mixed potential theory is established, providing clear stability boundaries for system parameters. Dynamic analyses theoretically verify system feasibility, and in-house hardware-in-the-loop experimental results fully validate the proposed HHESS along with the corresponding transient power allocation controls.
\end{abstract}
}


% \fi
\IEEEpeerreviewmaketitle

\vspace{-1em}
\section{Introduction}
% normal text (unk
%n if ever used by the IEEE):
% \vspace{-1em}
\subsection{Research Background}
With the deepening global energy shortage and escalating environmental concerns, developing clean alternative energy sources to replace conventional fossil fuels has become imperative. Among various emerging clean energy technologies, hydrogen, as a significant secondary energy carrier, exhibits substantial potential in building sustainable energy infrastructures due to its inherent cleanliness, efficiency, storability, and dispatchability\cite{Cha2025}\cite{IEAreport}. Rapid advancements in renewable generation and power electronics have increasingly promoted the formation of renewable-dominant power grids. By leveraging advanced power electronic equipment and sophisticated control technologies, these emerging power systems can efficiently integrate and manage multiple clean energy sources, substantially reducing traditional conversion losses and improving overall efficiency\cite{Ana2019}. Additionally, power-electronic-based renewable grids notably reduce carbon footprints and enhance energy utilization effectiveness.

However, due to the extensive replacement of traditional synchronous generators by power electronics equipment, the overall inertia of power system has substantially decreased\cite{Ben2022}, resulting in reduced dynamic frequency stability and increased vulnerability to external disturbances, which elevates the risk of loss of synchronization or even induces system collapse \cite{Hat2021, QingzuoMeng2025}. Moreover, such renewable-dominated grids are prone to issues such as harmonics, voltage fluctuations, and power disturbances, severely degrading power quality and posing significant challenges to grid security and stability\cite{Sta2022}. Therefore, effectively improving frequency stability and disturbance resisting capability in renewable-dominated power systems has become a critical research topic that calls for urgent solutions.

\vspace{-1em}
\subsection{Literature Review}\label{subsec:litreview}
Currently, studies over hydrogen-electric integrated systems combining hydrogen production and power grids has received wide attentions. In such systems, electrolyzers are employed to produce hydrogen serving as a clean energy carrier that can be subsequently utilized across various applications \cite{Md2023}. Nevertheless, these investigations focus more on maximizing the harvest of hydrogen, while few of them explores the flexibility that hydrogen production systems can provide to actively support grid frequency regulations \cite{Pan2020}.

In fact, hydrogen production systems can be crafted as adjustable loads and they would exhibit significant potential for participating in grid regulations. There are normally two types of electrolyzers out of the shelf that can be reshaped as flexible loads, which are proton exchange membrane electrolyzers (PEMELs) and alkaline electrolyzers (AELs), respectively. For the former, it is extensively known that PEMELs offer fast dynamic response (seconds), making them ideal for the applications with intermittent renewable energy. In \cite{Tav2023}, a PEMEL is integrated into power grids as flexible loads to effectively mitigate frequency fluctuations caused by renewable energy variations. Such integration enhances grid responsiveness to disturbances and supports the development of modern power systems. In \cite{Doz2021}, a dynamic modeling framework is proposed to evaluate the fast frequency response capabilities of utility-scale PEMELs, highlighting how their performance depends on system conditions and demonstrating their potential to enhance frequency resilience in low-inertia, renewable-rich power systems. In \cite{Hos2023}, rapid frequency response based on PEMELs is investigated, demonstrating the improved dynamic regulation ability and enhanced frequency stability of power grid under renewable-induced uncertainties. In \cite{Doz2023}, virtual synchronous machine (VSM) control is applied to PEMELs, enabling them to provide virtual inertia and frequency support. 

It is worth highlighting that PEMELs normally incur higher capital costs as their proper system functioning requires the use of precious metal catalysts. Their working efficiency is commonly moderate at 60-70\%. In contrast, AELs constructed based nickel-based electrodes would give much lower front costs; they are of higher operating efficiency at 70-80\%. For those costs-sensitive engineering practices, AELs apparently offer more viable options \cite{Gu2024}. In \cite{ChuanjunHuang2023}, AELs are adopted as dispatchable loads to alleviate the frequency regulation stress of bulk system subjected to high intermittent wind generations. The power reference for AELs consumption is proportional to the real-time differentiation of grid frequency. In \cite{QingchaoSong2024}, AELs are combined with several batteries to assist in the energy management of overall DC system, and the operating power of AELs could be consistently maintained above the minium recommended level. A multi-mode optimal electrolysis converter strategy is proposed in \cite{YanghongXia2024} such that the power of AELs can be better coordinated along with photovoltaic power fluctuations. Similar efforts can also be found in \cite{YanghongXia2024b} where direct current mode and pulse current mode are applied to AELs to boost the working efficiency in a wider range. In \cite{Xu2025}, a supercaptor is employed to compensate for the electrical characteristics of AELs and the holistic system performance can be escalted. By taking into account of various dynamic physical models, a centralized model predictive control is proposed in \cite{AoboGuan2025} to figure out optimal electrolysis power profiles in minute time scale.

\vspace{-1em}
\subsection{Research Gap and Contributions}
{\color{black}
Table~\ref{tab:lit_comparison} presents a comprehensive comparison between our work and the prior studies discussed in Section~\ref{subsec:litreview}. Most existing research has focused either on hydrogen-coupled system planning / optimization or on investigating single-technology electrolyzers for frequency supports. The hybrid system combining PEMEL, AEL, and supercapacitor (SC), wherein all of their advantages in dynamic response are properly coordinated, is underplayed. Besides, the relevant decentralized, time-scale–aware droop coordination is also absent, and the corresponding large-signal stability for electrolyzer based system is seldom.
\begin{table*}[t]
  \caption{\color{black} Comparisons Between the Proposed HHESS and Related Works}
  \label{tab:lit_comparison}
  \centering
  \setlength{\tabcolsep}{4pt}
  \renewcommand{\arraystretch}{1.15}
  {\color{black} % —— 表格文字整体设为红色 ——
  \begin{tabularx}{\textwidth}{@{}lCCCCC@{}}
    \toprule
    \textbf{References} &
    \makecell{\textbf{Hybrid electrolyzer}\\\textbf{architecture}} &
    \makecell{\textbf{Equipped with}\\\textbf{energy storage}} &
    \makecell{\textbf{Active grid}\\\textbf{frequency support}} &
    \makecell{\textbf{Decentralized dynamic}\\\textbf{power sharing control}} &
    \makecell{\textbf{Large-signal}\\\textbf{stability analysis}} \\
    \midrule
    {\cite{Pan2020}, \cite{YanghongXia2024}}   & \xmark & \xmark & \xmark & \xmark & \xmark  \\
    {\cite{QingchaoSong2024}}        & \xmark & \cmark & \xmark & \cmark & \xmark         \\
    {\cite{ChuanjunHuang2023}}        & \xmark & \xmark & \cmark & \cmark & \xmark          \\
    {\cite{Tav2023}, \cite{Doz2021}, \cite{Hos2023}}   & \xmark & \xmark & \cmark & \xmark & \xmark     \\
    {\cite{Doz2023}}        & \xmark & \xmark & \cmark & \xmark & \xmark          \\
    {\cite{Xu2025}}        & \xmark & \cmark & \cmark & \cmark & \xmark          \\
    \textbf{This paper} & \cmark & \cmark & \cmark & \cmark & \cmark \\
    \bottomrule
  \end{tabularx}
  } % —— 结束红色范围 ——
  \vspace{-1em} % 需要的话可微调与正文距离
\end{table*}

Motivated by these gaps, this paper proposes a hybrid hydrogen electrolyzer–supercapacitor system (HHESS) that incorporates SC alongside PEMEL and AEL to enhance both dynamic response and frequency support performance. Moreover, a novel transient power allocation control scheme for HHESS is also presented. When the HHESS is coupled to a power grid through a three phase AC/DC inverter (see Fig. \ref{Fig. Schematic diagram of HHESS}), the inverter helps to burst out transient power to actively offer inertia response service to the grid. The full time scale power flow over the inverter would be spontaneously decomposed into low-, mid-, and high- frequency components, which are then allocated to AEL, PEMEL, and SC, respectively. By doing so, not only can the HHESS naturally serve as a flexible load actively contributing inertia to the power grid, but the dynamic power responses of AEL, PEMEL, and SC would be well orchestrated, thus measurably prolonging the lifespan of overall HHESS. Key contributions are summarized as follows.}

1) Dynamic power allocation in HHESS and active inertia support to grid. The AEL is designated with a static V-P droop, whereas the PEMEL and the SC are assigned with the proposed dynamic integral droop (DID) and capacitive integral droop (CID), respectively. The power reference for the interfacing three phase inverter between HHESS and grid is generated by inertia control loop and the HHESS can then spontaneously emulate inertia to the grid. Note that the inverter does not generate power itself. It just works as a bridging device that drags inertia from HHESS giving to the grid and transfer static power from grid to feed HHESS. Under grid frequency disturbances, the inverter dynamically adjusts output power, allocating mid -frequency power components autonomously to PEMEL, and assigning low-frequency power to AEL. SCs covers high-frequency power and provide immediate transient power support, alleviating dynamic operational stresses on PEMEL and AEL. When transient processes concludes, the SC power gradually reduces to zero, leaving itself idle until the next grid disturbance occurs. In steady-state conditions, PEMEL and AEL fully account for the load demands, thus preventing frequent discharging of SCs and prolonging the over operational life of HHESS.

% In transient state, by combining V-P droop, DID, and CID, the overall load demand of HHESS can be autonomously split into into low-, middle-, and high- frequency components which are covered by AEL, PEMEL, and SC, respectively. In steady state, AEL and PEMEL consume power at a specific level while SC power gradually reduces to zero. Thanks to coordinated power allocation in HHESS, 

 % considering significant differences in dynamic characteristics among components. 

2) Autonomous state-of-charge (SOC) recovery of SC. Thanks to the pretense of CID, the designed control scheme not only dynamically coordinates power allocation among different system components but also autonomously restores SC SOC without requiring additional communication. Following transient power support provided by SC, the energy released spontaneously returns to the SC itself. Throughout the transient process, no energy from SC is transferred externally, allowing SC SOC to remain autonomously at original levels. This ensures stable long-term operation, completely avoids SOC depletion issues, and significantly extends the SC expectation.

%3)	Flexible expansion capability of HHESS. To enhance system flexibility in engineering applications, a detailed system expansion methodology is introduced. The method clearly describes how to effectively increase the quantity and capacity of PEMEL, AEL, and SC units without substantially altering the existing control framework. By adopting these guidelines, the system can be flexibly adjusted and expanded based on actual operating conditions and demands, ensuring consistent system functionality and stability while reducing the complexity and costs associated with system expansion.

3) Large-Signal stability analysis and parameter design. Mixed potential theory (MPT), a large-signal mathematical model for the proposed HHESS regulated by transient power allocation control is established. Stability criteria under significant disturbance conditions are rigorously derived, and clear stability boundaries for key system parameters are defined. This stability analysis method provides robust theoretical foundations for further optimization of control strategies and rational design of system parameters, ensuring excellent dynamic stability performance under diverse and complex real-world conditions.

The remainder of this paper is organized as follows. Section II details the basic configurations and mathematical modeling of the proposed HHESS. Section III elaborates on the proposed transient power allocation scheme within HHESS where novel DID and CID are elaborated. The active inertia response of HHESS and the dynamic power coordination among AEL, PEMEL, and SC, are also meticulously analyzed. Section IV elaborates on the modular scalability and decentralized control strategy for system expansion. Section V evaluation the whole system stability under large disturbances based on MPT. The effectiveness of the proposed system structure, control strategies, and stability analyses would be verified by hardware-in-the-loop (HIL) experiments in Section VI. Finally, conclusions are drawn in Section VII.

\vspace{-0.5em}
\section{The Proposed Hybrid Hydrogen Electrolyzer-Supercapacitor System (HHESS)}
An electrolyzer is a device that utilizes external electrical energy to drive chemical reactions. In hydrogen electrolyzers, the fundamental reaction involves decomposing water into hydrogen and oxygen, with the general reaction presented as follows \eqref{eq:overall_electrolysis}:
\begin{equation}
\setlength{\abovedisplayskip}{3pt}
\setlength{\belowdisplayskip}{3pt}
\ce{H2O ->[\text{Energy}] H2 + \frac{1}{2}O2}.
\label{eq:overall_electrolysis}
\end{equation}

During this process, input electrical energy is converted into chemical energy and stored in the produced hydrogen. Currently, two predominant electrolyzer technologies, namely PEMEL and AEL, are widely adopted. Both consist of an anode and a cathode, separated by an electrolyte facilitating ionic conduction\cite{Moh2024}. 

The AEL adopts a potassium hydroxide (KOH) aqueous solution as the electrolyte, where hydroxide ions ($\mathrm{OH}^-$) serve as charge carriers between electrodes. The alkaline environment facilitates ionic conduction; however, the transport rate is constrained by factors such as electrolyte concentration, temperature, and diffusion. As a result, AELs exhibit limited ion mobility under rapid dynamic conditions. Furthermore, due to high activation energy barriers for hydrogen and oxygen evolution reactions, their electrochemical efficiency—especially at high current densities—remains low. To mitigate these issues, AELs are typically designed with large electrode areas and operate at lower current densities\cite{Sed2024}. Although this design reduces cost, it introduces significant system inertia and slow response, as the liquid electrolyte requires time to stabilize. Consequently, AEL performance is restricted in applications demanding fast power modulation.

In contrast, PEMEL utilizes a solid polymer membrane as the electrolyte, with protons ($\mathrm{H}^+$) serving as charge carriers in an acidic environment. Proton conduction occurs directly across the membrane, minimizing resistance and enabling high reaction efficiency even at elevated current densities. This allows PEMEL to deliver rapid dynamic response and high power density. However, these advantages come at the cost of increased manufacturing complexity and reduced operational lifespan. Key components—such as the proton exchange membrane and catalysts—require advanced materials and suffer degradation under sustained acidic exposure, leading to corrosion and catalyst deactivation\cite{Nor2024}. These challenges limit the economic viability and scalability of PEMEL in industrial applications. Technical and economic parameters of both electrolyzer technologies are summarized in Table~\ref{tab:PEMEL_AEL}\cite{Sub2023b}.

%The simplified equivalent circuit of electrolyzers is depicted in Fig.~\ref{Fig. Simplified Equivalent Circuit of the Electrolytic Cell}.
\begin{comment}
\begin{figure}[t]
	\setlength{\abovecaptionskip}{-3pt}
	\setlength{\belowcaptionskip}{-5pt}
	\centering
	\includegraphics[scale=1]{1.png}
	\caption{Simplified Equivalent Circuit of the Electrolytic Cell.}
	\label{Fig. Simplified Equivalent Circuit of the Electrolytic Cell}
	\vspace{-1em}
\end{figure} 
\end{comment}

\begin{figure}[t]
	\setlength{\abovecaptionskip}{-3pt}
	\setlength{\belowcaptionskip}{-5pt}
	\centering
	\includegraphics[scale=1]{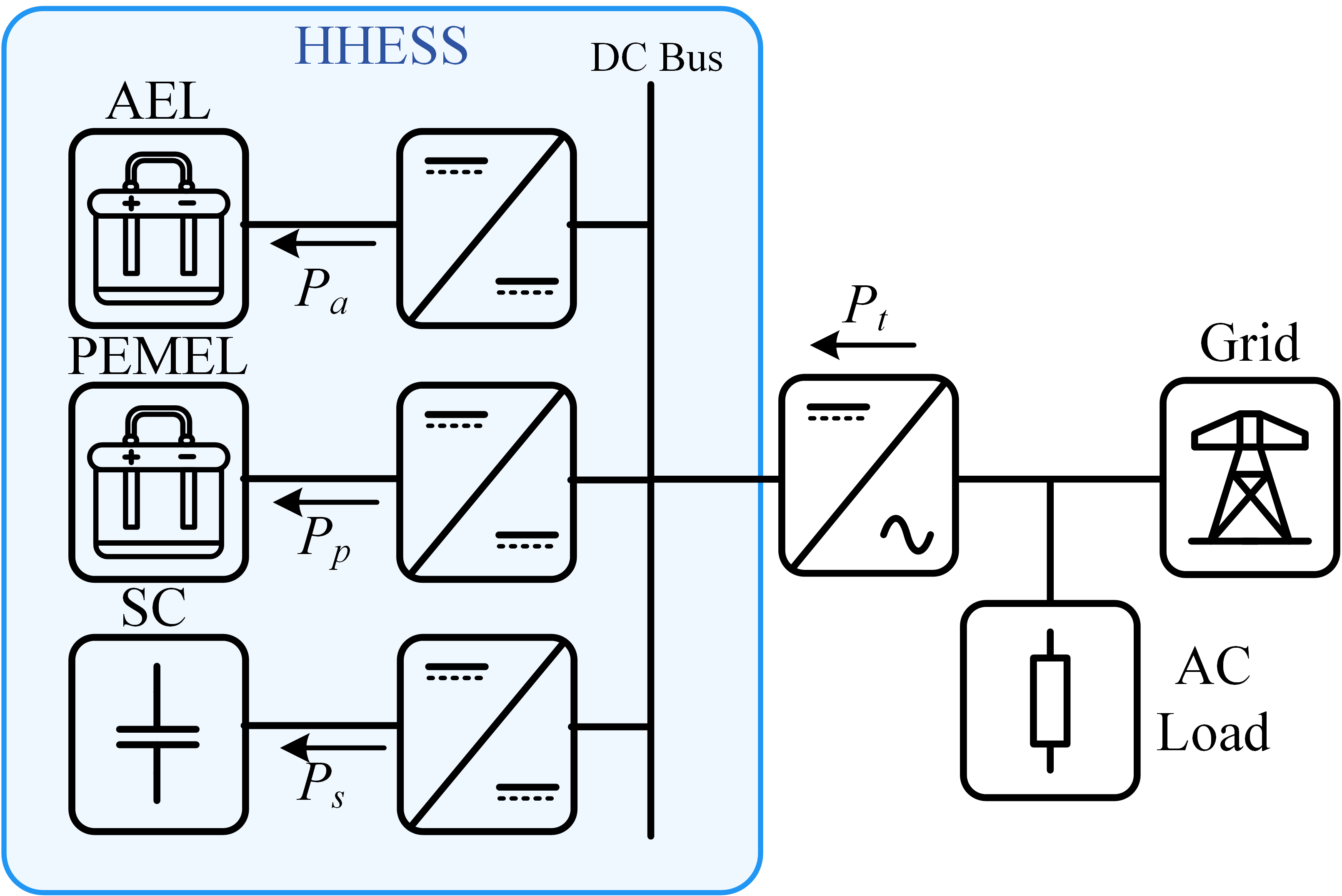}
	\caption{Schematic diagram of the proposed HHESS.}
	\label{Fig. Schematic diagram of HHESS}
	\vspace{-1em}
\end{figure}

\begin{comment}
\begin{table}[htbp]
    \vspace{-1em}  % 缩小表格与上文间距
    \caption{Economic and Technical Parameters of PEMEL and AEL}
    \label{tab:PEMEL_AEL}
    \setlength{\abovecaptionskip}{-5pt}
    \setlength{\belowcaptionskip}{-10pt}
    \vspace{-1em}  % 缩小标题与表格间距
    \centering
    \begin{tabular}{lcc}
        \toprule
        \textbf{Parameter} & \textbf{PEMEL} & \textbf{AEL} \\
        \midrule
        Capital cost (CNY/kW) & 12{,}000 & 4{,}000 \\
        Annual O\&M cost (CNY/kW) & 480 & 160 \\
        System lifetime (khr) & 15--80 & 90--120 \\
        Rated current density (A/cm\textsuperscript{2}) & 4--6 & 2 \\
        Load range (\%) & 0--160 & 15--100 \\
        Power ramping rate & 40\% /s & 10\% /s \\
        \bottomrule
    \end{tabular}
    \vspace{-1em}  
\end{table}
\end{comment}

\begin{table}[htbp]
    \vspace{-1em}  % 缩小表格与上文间距
    \caption{{\color{black}Key Economic and Technical Parameters of PEMEL and AEL~\cite{Sub2023b}}}
    \label{tab:PEMEL_AEL}
    \setlength{\abovecaptionskip}{-5pt}
    \setlength{\belowcaptionskip}{-10pt}
    \vspace{-1em}  % 缩小标题与表格间距
    \centering
    {\color{black}
    \begin{tabular}{c c c}
        \toprule
        \textbf{Parameter} & \textbf{PEMEL} & \textbf{AEL} \\
        \midrule
        Total stack cost (€/kW) & 384--1071 & 242--388 \\
        Direct cost (€/kW) & 308--332 & 192--205 \\
        Material cost only (€/kW) & 190 & 118 \\
        Power density (W/cm\textsuperscript{2}) & 4.5 & 0.5 \\
        Current density (A/cm\textsuperscript{2}) & 2.0 & 0.245 \\
        Operating pressure (bar) & 20 & Ambient \\
        Power ramping rate & Fast & Slow \\
        Technology maturity (TRL) & 6--7 & 8--9 \\
        System lifetime (khr) & 15--80 & 90--120 \\
        \bottomrule
    \end{tabular}
    }
    \vspace{-1em}  
\end{table}

\begin{comment}
\begin{table}[htbp]
    \vspace{-1em}
    \caption{Economic and Technical Parameters of PEMEL and AEL (Baseline 2020)}
    \label{tab:PEMEL_AEL_2020}
    \setlength{\abovecaptionskip}{-5pt}
    \setlength{\belowcaptionskip}{-10pt}
    \vspace{-1em}
    \centering
    \begin{tabular}{lcc}
        \toprule
        \textbf{Parameter} & \textbf{PEMEL (Baseline)} & \textbf{AEL (Baseline)} \\
        \midrule
        Total stack cost (€/kW) & 384--1071 & 242--388 \\
        Direct cost (Material + Labor + Manufacturing) (€/kW) & 308--332 & 192--205 \\
        Material cost only (€/kW) & 190 & 118 \\
        Power density (W/cm\textsuperscript{2}) & 4.5 & 0.5 \\
        Current density (A/cm\textsuperscript{2}) & 2.0 & 0.245 \\
        Operating pressure (bar) & 20 & Ambient \\
        Temperature at nominal load (°C) & 55 & 80 \\
        Stack size (MW) & 0.67 & 2.2 \\
        Active surface area (m\textsuperscript{2}) & 0.10 & 2.1 \\
        Voltage (V) & 2.0 & 1.85 \\
        Ramp response & Fast & Slow \\
        Critical material dependence & Pt, Ir, Au, Ti & Ni \\
        Technology maturity (TRL) & 6--7 & 8--9 \\
        \bottomrule
    \end{tabular}
    \vspace{-1em}
\end{table}    
\end{comment}

{\color{black}
AEL and PEMEL technologies present distinctive advantages and limitations: AEL, with lower costs and longer lifespan, is prevalent in industrial scenarios but suffers from slower dynamic response; conversely, PEMEL demonstrates superior dynamic performance and higher power density, thus excelling in applications requiring high responsiveness, but its broad adoption is hindered by its high cost and limited operational lifespan. To overcome these limitations of individual electrolyzer technologies, and to meet economic demands while enhancing rapid responsiveness to grid frequency adjustments, a hybrid system integrating both electrolyzers is necessary. Such a hybrid system effectively combines the fast response of PEMEL, the economic advantages of AEL, and SC capability to provide instantaneous power support, thus mitigating response pressures on electrolyzers and extending overall system lifespan. The schematic configuration of the proposed HHESS is illustrated in Fig.~\ref{Fig. Schematic diagram of HHESS}.
}

This study proposes the HHESS employing hierarchical control for dynamic power sharing coordination. Under steady-state conditions, both AEL and PEMEL simultaneously produce hydrogen. When grid disturbances occur, SCs provide instantaneous power support through their rapid charge-discharge capability, allowing PEMEL to swiftly respond to power variations, followed by slower power adjustments from the AEL. Such multi-timescale coordination effectively utilizes SC transient response and PEMEL rapid adjustment capabilities, thereby stabilizing grid frequency and simultaneously enhancing both dynamic performance and economic efficiency of the system.

\vspace{-0.5em}
\section{The Proposed Control Scheme for HHESS} %Section II
\label{sec:control}
\begin{figure*}[t]
	\setlength{\abovecaptionskip}{-3pt}
	\setlength{\belowcaptionskip}{-5pt}
	\centering
	\includegraphics[scale=1]{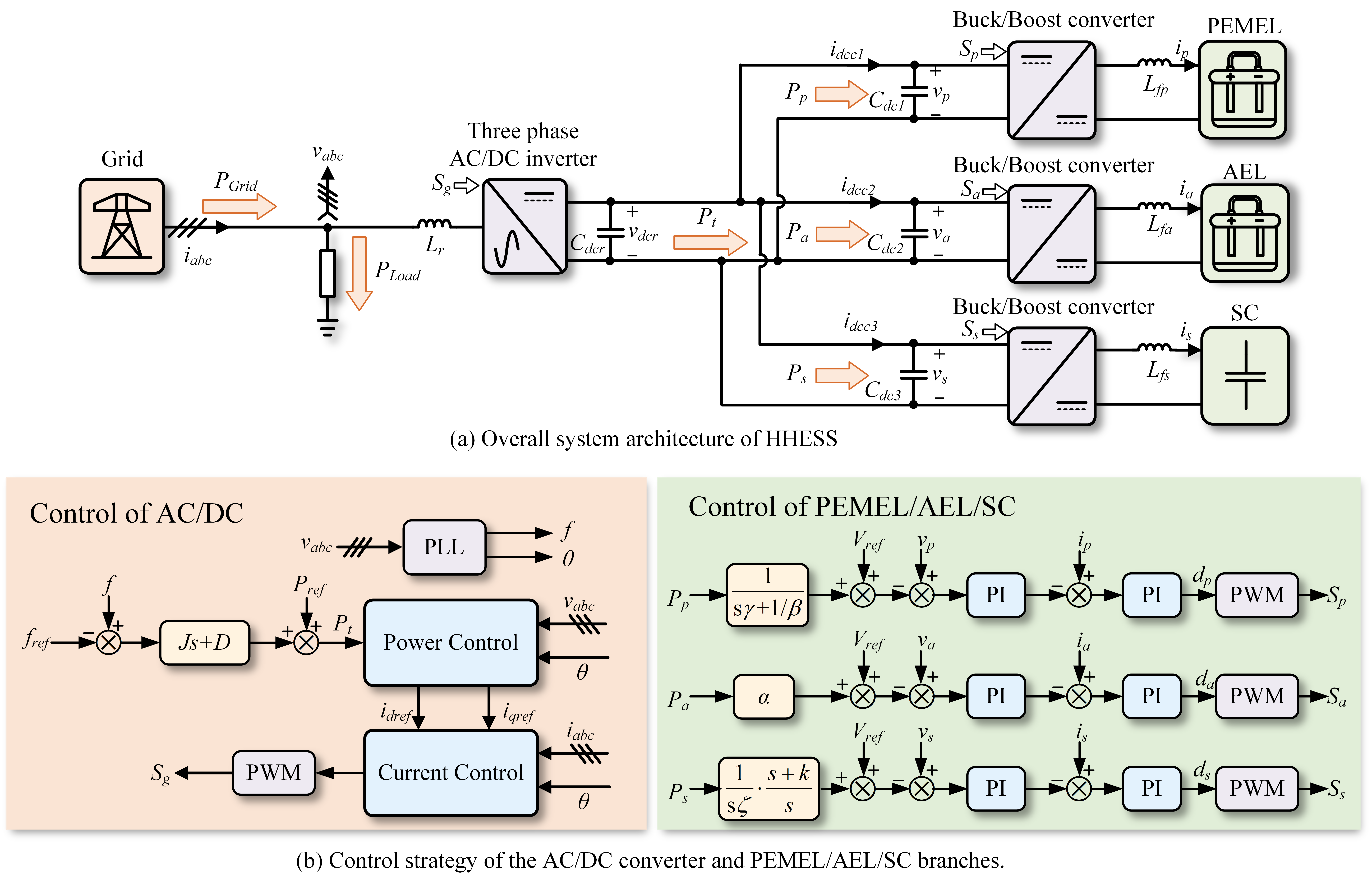}
	\caption{\textcolor{black}{System design and proposed control strategy for HHESS.}}
	\label{Fig. 8}
	\vspace{-1em}
\end{figure*}
The proposed control strategy for HHESS is illustrated in Fig.~\ref{Fig. 8}. In this scheme, PEMEL, AEL, and SC branches are coordinated to achieve multi-timescale power sharing. High-frequency transient power components are autonomously handled by SC, middle-frequency power components are regulated by PEMEL, and low-frequency steady-state power is addressed by AEL. This coordination alleviates the dynamic burden on the electrolyzers. Additionally, a grid-connected inverter operating under active inertia provision control dynamically adjusts system power in response to frequency deviations. All converters operate independently without communication, supporting high reliability and modular scalability. The control mechanisms are described in detail in the following subsections.

\vspace{-1em}
\subsection{Transient Power Allocation Control}
\label{hhesscontrol}
In this subsection, a transient power allocation control strategy is developed for the HHESS. The proposed scheme employs three distinct droop control methods to coordinate the dynamic power sharing among PEMEL, AEL, and SC, enabling fast and slow response channels for system power fluctuations.

In the control system, AEL adopts a static V-P droop control strategy, which primarily regulates power output under steady-state conditions. This approach enables AEL to participate in system-level power allocation by adjusting its terminal voltage in response to its output power, thus contributing to coordinated power sharing:
\begin{equation}
\setlength{\abovedisplayskip}{3pt}
\setlength{\belowdisplayskip}{3pt}
v_a = V_{\mathrm{ref}} + \alpha P_a
\label{eq:AEL_droop}
\end{equation}
where $v_a$ and $P_a$ represent the DC bus voltage and the power input to the AEL, $V_{\mathrm{ref}}$ denotes the reference voltage of the DC bus, and $\alpha$ is the droop coefficient.

% This droop design takes into account the slower dynamic response of AEL. It provides accurate current sharing and favorable long-term economic operation in steady-state. Moreover, 
By adjusting the $\alpha$, the power sharing can be flexibly coordinated according to the rated power capacity of AEL:
\begin{equation}
\setlength{\abovedisplayskip}{3pt}
\setlength{\belowdisplayskip}{3pt}
\alpha = \frac{\Delta V_{\max}}{P_{a\mathrm{max}}}
\label{eq:AEL_alpha}
\end{equation}
where $\Delta V_{\max}$ denotes the maximum allowable voltage deviation of the DC bus, and $P_{a\mathrm{max}}$ is the rated steady-state power.

For PEMEL, a dynamic integral droop (DID) control strategy is employed, which combines capacitive and resistive dynamic characteristics to enable effective power sharing under both transient and steady-state conditions. The proposed DID can be expressed below,
\begin{equation}
\setlength{\abovedisplayskip}{3pt}
\setlength{\belowdisplayskip}{3pt}
v_p = V_{\mathrm{ref}} + \frac{1}{s \gamma + \frac{1}{\beta}} P_p.
\label{eq:PEMEL_droop}
\end{equation}

This droop control scheme takes advantage of PEMEL’s fast dynamic response. The incorporation of fast dynamic components significantly enhances the system’s capability to handle transient power fluctuations, compensating for the dynamic limitations of AEL.

The SC employs a capacitive integral droop (CID) control strategy, enabling it to participate in the overall power allocation through its inherent capacitive characteristics. The proposed CID can be expressed below,
\begin{equation}
\setlength{\abovedisplayskip}{3pt}
\setlength{\belowdisplayskip}{3pt}
v_s = V_{\mathrm{ref}} + \frac{1}{s \zeta} \cdot \frac{s + k}{s} P_s.
\label{eq:SC_droop}
\end{equation}

This control design fully utilizes the rapid charge and discharge capability of the SC, allowing the system to respond swiftly to instantaneous power disturbances.

{\color{black} Assume that the total power delivered to the DC bus from the grid is $P_t$, while $P_a$, $P_p$, and $P_s$ denote the DC power flowing into the DC/DC converters of AEL, PEMEL, and SC, respectively. Since the AEL, PEMEL, and SC branches are connected in parallel to the same DC bus, they share a common bus voltage $v_{dc}$. With these definitions, the following assumptions are adopted:
\begin{equation}
\setlength{\abovedisplayskip}{3pt}
\setlength{\belowdisplayskip}{3pt}
\left\{
\begin{aligned}
v_a &= v_p = v_s = v_{dc}\\
P_t &= P_a + P_p + P_s.
\end{aligned}
\right.
\label{eq:assum}
\end{equation}

Combining \eqref{eq:AEL_droop}, \eqref{eq:PEMEL_droop}, \eqref{eq:SC_droop} and \eqref{eq:assum}, the individual branch power can be derived as:

\begin{equation}
\setlength{\abovedisplayskip}{3pt}
\setlength{\belowdisplayskip}{3pt}
P_a = \frac{(s + k)/\alpha}{s^2 (\zeta + \gamma) + s \left( k \gamma + \frac{1}{\alpha} + \frac{1}{\beta} \right) + k \left( \frac{1}{\alpha} + \frac{1}{\beta} \right)}P_{t}
\label{eq:Ia}
\end{equation}

\begin{equation}
\setlength{\abovedisplayskip}{3pt}
\setlength{\belowdisplayskip}{3pt}
P_p = \frac{(s + k)(s \gamma + 1/\beta)}{s^2 (\zeta + \gamma) + s \left( k \gamma + \frac{1}{\alpha} + \frac{1}{\beta} \right) + k \left( \frac{1}{\alpha} + \frac{1}{\beta} \right)}P_{t}
\label{eq:Ip}
\end{equation}

\begin{equation}
\setlength{\abovedisplayskip}{3pt}
\setlength{\belowdisplayskip}{3pt}
P_s = \frac{s^2 \zeta}{s^2 (\zeta + \gamma) + s \left( k \gamma + \frac{1}{\alpha} + \frac{1}{\beta} \right) + k \left( \frac{1}{\alpha} + \frac{1}{\beta} \right)}P_{t}.
\label{eq:Is}
\end{equation}
}

{\color{black}
As shown in \eqref{eq:Ia}--\eqref{eq:Is}, the individual branch powers $P_a$, $P_p$, and $P_s$ can be expressed as the outputs of second-order filters applied to the total power $P_t$, with all filters sharing a common denominator. By comparing this denominator with the canonical second-order system form, the following relationships are obtained:
\begin{equation}
\setlength{\abovedisplayskip}{3pt}
\setlength{\belowdisplayskip}{3pt}
s^2 \;+\; \frac{k\gamma + \tfrac{1}{\alpha} + \tfrac{1}{\beta}}{\zeta + \gamma}\,s
\;+\; \frac{k\!\left(\tfrac{1}{\alpha} + \tfrac{1}{\beta}\right)}{\zeta + \gamma}
\;=\; s^2 + 2\xi \omega_0 s + \omega_0^2
\label{eq:second_order_compare}
\end{equation}
where $\omega_0$ is the natural frequency and $\xi$ is the damping ratio, leading to:
\begin{equation}
\setlength{\abovedisplayskip}{3pt}
\setlength{\belowdisplayskip}{3pt}
\left\{
\begin{aligned}
\omega_0 &= \sqrt{ \frac{k\left( \tfrac{1}{\alpha} + \tfrac{1}{\beta} \right) }{\zeta + \gamma} } \\
\xi &= \frac{ k\gamma + \tfrac{1}{\alpha} + \tfrac{1}{\beta} }{ 2\sqrt{ (\zeta + \gamma) \, k \left( \tfrac{1}{\alpha} + \tfrac{1}{\beta} \right) } }.
\end{aligned}
\right.
\label{eq:omega_xi_basic}
\end{equation}

Define the practical sharing indices \(k_1 = \alpha/\beta\) and \(k_2 = \zeta/\gamma\), where \(k_1\) represents the steady-state power-sharing ratio between the PEMEL and AEL branches, and \(k_2\) quantifies the power-sharing ratio between the SC and PEMEL branches in the transient state. In a practical HHESS implementation, \(k_1\) and \(k_2\) are typically chosen to fulfill the desired power allocation requirements across PEMEL, AEL, and SC. Substituting these expressions into \eqref{eq:omega_xi_basic} yields:
\begin{equation}
\setlength{\abovedisplayskip}{3pt}
\setlength{\belowdisplayskip}{3pt}
\left\{
\begin{aligned}
\omega_0&=\sqrt{\frac{k\,\tfrac{1+k_1}{\alpha}}{\gamma(1+k_2)}} \,\\[3pt]
\xi &= \frac{k\gamma + \tfrac{1+k_1}{\alpha}}{2\sqrt{\gamma(1+k_2)\,k\,\tfrac{1+k_1}{\alpha}}}\, .
\end{aligned}
\right.
\label{eq:omega_xi_k1k2}
\end{equation}

The relationship between $\omega_0$ and the cutoff frequency $\omega_c$ is taken as\cite{Qianwenxu2017}
\begin{equation}
\setlength{\abovedisplayskip}{3pt}
\setlength{\belowdisplayskip}{3pt}
\omega_0 \;=\; \frac{\omega_c}{\sqrt{\,1+2\xi^2 \;+\; \sqrt{\,1+2\xi^2\,} \;+\; 1\,}} 
\label{eq:omega0_from_omegac}
\end{equation}
with $\omega_c$ set by a design time constant $\tau$ as
\begin{equation}
\setlength{\abovedisplayskip}{3pt}
\setlength{\belowdisplayskip}{3pt}
\omega_c \;=\; \frac{1}{\tau}.
\label{eq:omega_c_from_tau}
\end{equation}

The time constant $\tau$ can be adjusted in accordance with manufacturer specifications so as to respect the inherently slow dynamics of the AEL, while the PEMEL and SC controllers are employed to enhance the overall system response speed. Accordingly, for a specified $\tau$ and damping ratio $\xi$, and for chosen power-sharing indices $k_1$ and $k_2$, the parameters $\gamma$ and $k$ can be determined from \eqref{eq:omega_xi_k1k2}–\eqref{eq:omega_c_from_tau} as:
\begin{equation}
\setlength{\abovedisplayskip}{3pt}
\setlength{\belowdisplayskip}{3pt}
\left\{
\begin{aligned}
\gamma &= \frac{1+k_1}{\alpha\,\omega_0}\left(\xi - \sqrt{\,\xi^2 - \frac{1}{1+k_2}}\,\right) \\
k &= \frac{\omega_0^2\,\alpha\,\gamma\,(1+k_2)}{1+k_1}.
\end{aligned}
\right.
\label{eq:gamma_k_from_tau}
\end{equation}
}
\begin{figure}[t]
	\setlength{\abovecaptionskip}{-3pt}
	\setlength{\belowcaptionskip}{-5pt}
	\centering
	\includegraphics[scale=1]{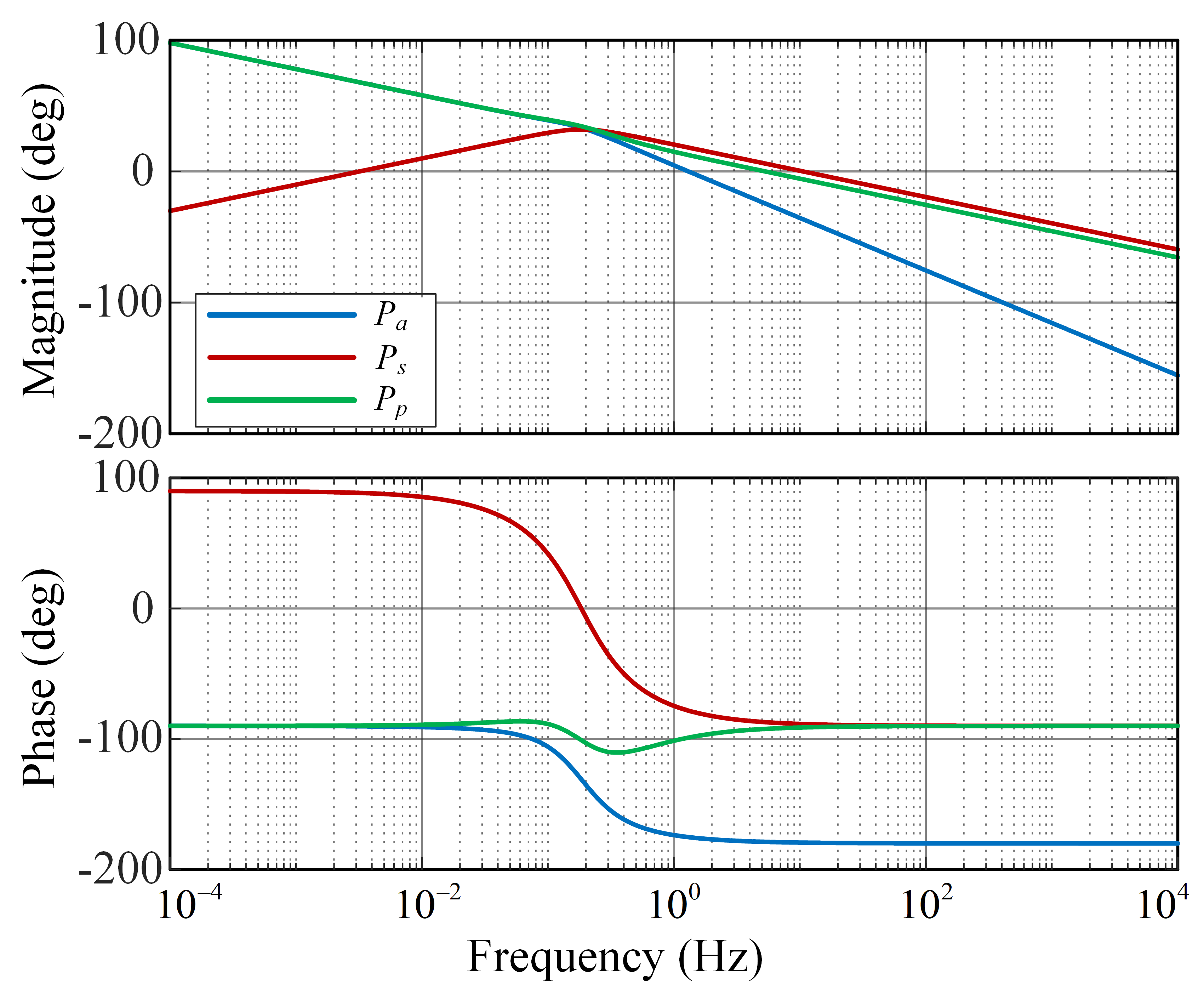}
	\caption{The dynamic power responses of PEMEL, AEL and SC in frequency domain.}
	\label{Fig. 3}
	\vspace{-1em}
\end{figure}
\begin{figure}[t]
	\setlength{\abovecaptionskip}{-3pt}
	\setlength{\belowcaptionskip}{-5pt}
	\centering
	\includegraphics[scale=1]{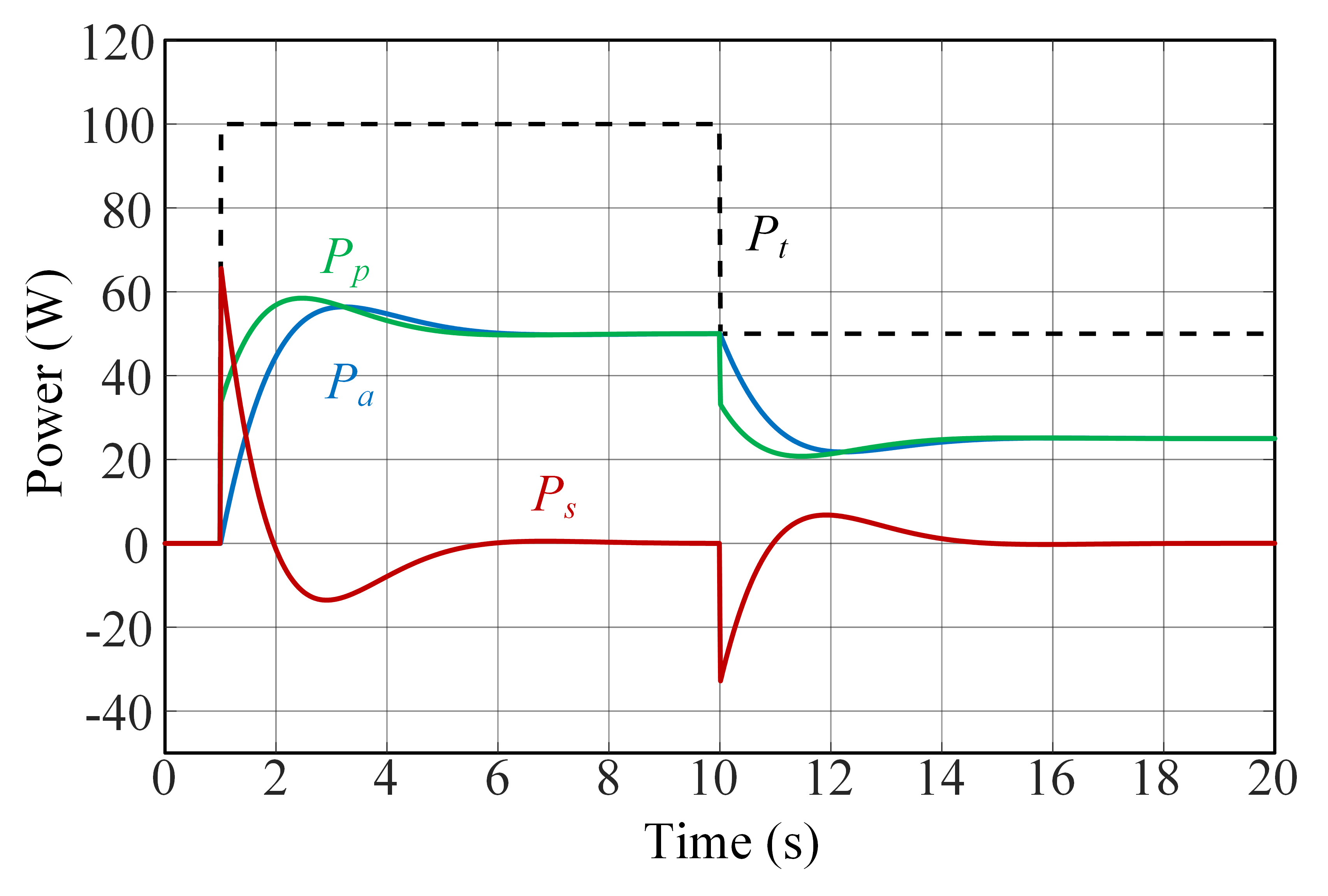}
	\caption{The dynamic power responses of PEMEL, AEL and SC in time domain.}
	\label{Fig. 4}
	\vspace{-1em}
\end{figure}
{\color{black}
\begin{figure}[t]
	\setlength{\abovecaptionskip}{-3pt}
	\setlength{\belowcaptionskip}{-5pt}
	\centering
	\includegraphics[scale=1]{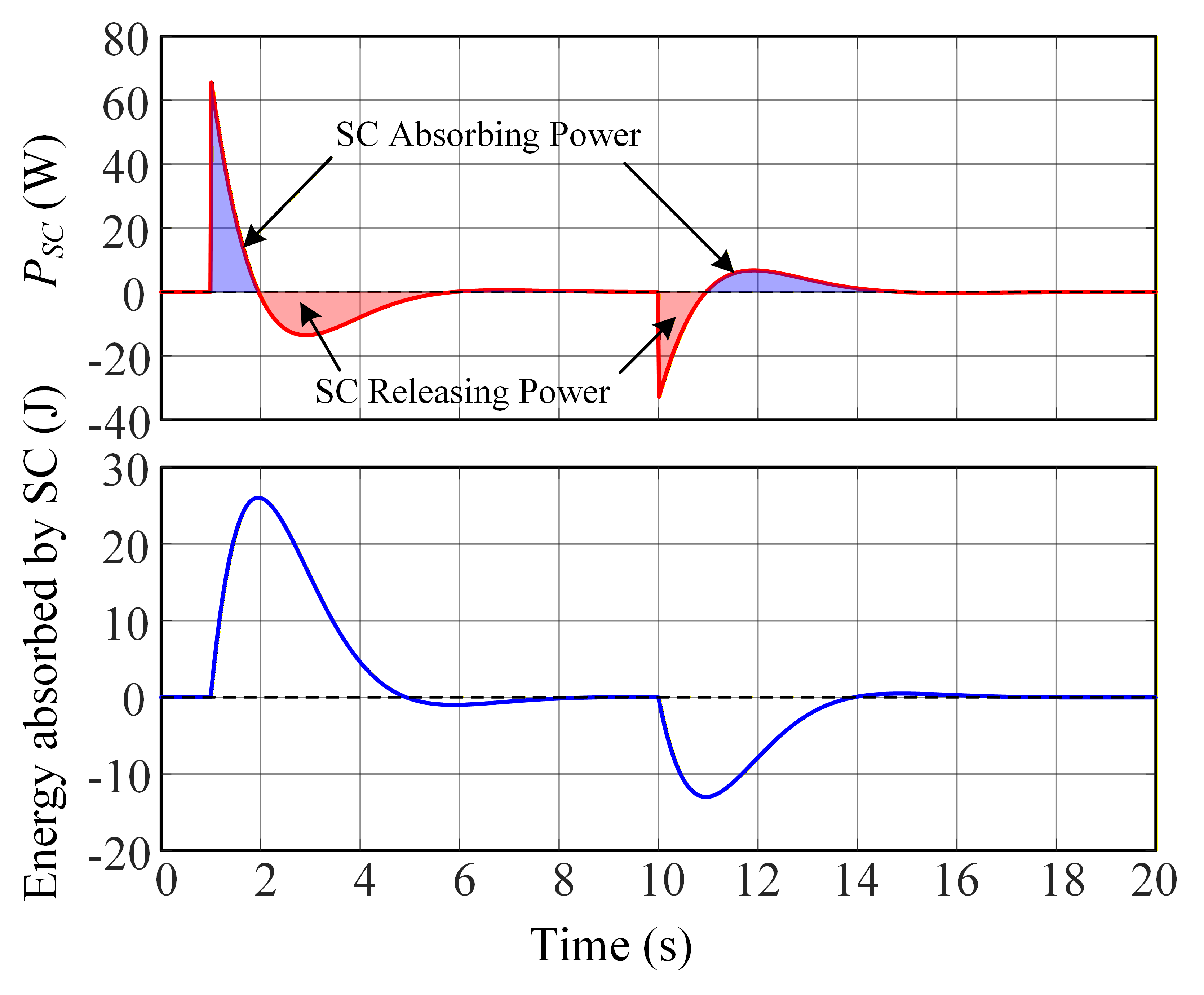}
	\caption{\textcolor{black}{Dynamic responses of $P_\mathrm{SC}$ and SC SOC. Durint 0s-6s, the power released by SC is absorbed by AEL+PEMEL.}}
	\label{Fig. 5}
	\vspace{-1em}
\end{figure}
}

{\color{black}
To investigate the dynamic power allocation characteristics of the system, the control parameters are set as $\alpha = \beta = 1/200$ V/W, $\gamma = 100$ W·s/V, $\zeta = 1/50$ W·s/V, and $k = 1$ s$^{-1}$. Based on the derived power transfer functions, the Bode plots of the three power branches are illustrated in Fig.~\ref{Fig. 3}. }

As observed from Fig.~\ref{Fig. 3}, the proposed droop-based control mechanism effectively decomposes the total system power into distinct frequency components. Specifically, the PEMEL and the SC, both characterized by fast dynamic responses, are primarily responsible for handling transient power fluctuations and regulating high-frequency components. In contrast, the AEL, due to its relatively slower dynamic behavior, compensates for the low-frequency components. This dynamic power allocation strategy ensures that each device operates within its optimal dynamic bandwidth, thereby improving system stability and enhancing transient response performance.

{\color{black}
With the control parameters defined above, the total power $P_t$ delivered from the DC bus to the PEMEL, AEL, and SC branches is programmed to step from $0\mathrm{W}$ to $100\mathrm{W}$ at $t$=1s, and then from $100\mathrm{W}$ to $50\mathrm{W}$ at $t$=10s.  The resulting power trajectories are depicted in Fig.~\ref{Fig. 4}: when a power disturbance occurs, the SC first handles the high-frequency power surge, the PEMEL regulates the middle-frequency component, and the AEL subsequently addresses the low-frequency component. Once steady state is reached, the powers of the PEMEL and AEL converge, while the SC power naturally returns to zero.
}

The SOC of the SC is likewise governed by the proposed CID as introduced earlier, and its power response is described by \eqref{eq:Is}. By converting the time-domain energy variation $\Delta Q_{\mathrm{SC}}$ to its Laplace-domain limit form, one obtains the SC energy change before and after the power response:
\begin{equation}
\setlength{\abovedisplayskip}{3pt}
\setlength{\belowdisplayskip}{3pt}
\Delta Q_\mathrm{SC} = \int_0^\infty P_{s}(t) dt = \lim_{s \to 0} P_{s}(s) = 0.
\label{eq:Delta_Q_SC}
\end{equation}

The corresponding SOC expression is given in \eqref{eq:SOC_SC} , where $SOC_{\mathrm{SC}}$ represents the current state of charge, $\mathrm{SOC}_{\mathrm{SC0}}$ is the initial SOC, $P_{\mathrm{s}}$ is the SC power and $E_{\mathrm{SC}}^{\mathrm{rated}}$ is the rated energy capacity of the SC.
\begin{equation}
\setlength{\abovedisplayskip}{3pt}
\setlength{\belowdisplayskip}{3pt}
\mathrm{SOC}_{\mathrm{SC}} = \mathrm{SOC}_{\mathrm{SC0}} + \int_0^t \frac{P_{s}}{E_{SC}^{rate}} dt.
\label{eq:SOC_SC}
\end{equation}

{\color{black}
Combining \eqref{eq:Delta_Q_SC} and \eqref{eq:SOC_SC} leads to $\mathrm{SOC}_{\mathrm{SC}} = \mathrm{SOC}_{\mathrm{SC}0}$, indicating that the SC branch maintains its original SOC throughout the entire power-response process. Fig.~\ref{Fig. 5} illustrates the SC power response in Fig.~\ref{Fig. 4} together with the energy released and re-absorbed by the SC. After providing transient support, the released energy spontaneously flows back into the SC, completing the SOC self-recovery. 

It should be noted that the SOC recovery mechanism presented in this paper is based on the presumption that there is no power loss in power conversion stage. This might not always hold in practice. When the case that converter efficiency is not high or SC  equivalent series resistance is taken into consideration, the integral of $P_s$ in (\ref{eq:SOC_SC}) is non-zero. Then, the SOC of SC would gradually decrease along with time elapse. In spite of the weakness, the proposed SOC recovery control scheme still keeps SC operating longer than that in system with SOC restoration control absent. Implementing SOC recovery control enables the SC to withstand up to ten times more charge–discharge cycles compared to an SC without SOC recovery. This advantage substantiates the superiority of the proposed control framework.  
}

Consequently, the proposed control strategy not only achieves dynamic power coordination among the PEMEL, AEL, and SC within the HHESS, but also enables autonomous SC SOC restoration without any additional communication overhead. Throughout the transient interval no SC energy is delivered to external loads; the SOC naturally returns to its initial level, ensuring long-term stable operation, preventing SOC depletion, and significantly extending SC service life.

{\color{black}
In summary, in the proposed HHESS, the DC bus voltage is cooperatively regulated by all droop-controlled branches. Each DC/DC converter adopts a dual-loop PI control structure to track the voltage reference generated by its corresponding droop controller. This allows for fast adjustment of $v_{\mathrm{dc}}$ following disturbances. Since the AEL, PEMEL, and SC units are connected in parallel, they share the same bus voltage $v_{\mathrm{dc}}$. This dual-loop structure also generates the PWM signals that drive the converters, thereby achieving dynamic power sharing among the HHESS modules while ensuring fast and stable system response.
}
\vspace{-1em}
\subsection{Autonomous Inertia Response}
%\lipsum[2-100]
This section investigates a control strategy aimed at fast frequency regulation in response to grid disturbances.

The frequency of low-inertia power systems is highly susceptible to external load fluctuations. When a sudden load change occurs, it may induce a frequency deviation which, if not properly and promptly mitigated, can escalate the operational risk of the grid. To address this issue, a fast frequency regulation method based on inverter-side power modulation is proposed. Specifically, when the grid frequency deviates from its nominal value, the inverter actively reduces the power drawn from the HHESS. This rapid power reduction alleviates the stress on the grid, suppresses frequency deviation, and ultimately enhances the stability and reliability of the overall power system\cite{Tor2014}.

The inverter control adopts an inertia emulation control strategy, and its control block diagram is illustrated in Fig.~\ref{Fig. 8}(b)\cite{Lin2019}. As shown in figure, the power regulation mechanism can be analytically described by the following equations, which govern the dynamic power output delivered from the inverter to the hydrogen production system:
\begin{equation}
\setlength{\abovedisplayskip}{3pt}
\setlength{\belowdisplayskip}{3pt}
\Delta f = \left( - P_{ref} + P_{t} \right) \cdot \frac{1}{Js + D}
\label{eq:freq_response}
\end{equation}
\begin{equation}
\setlength{\abovedisplayskip}{3pt}
\setlength{\belowdisplayskip}{3pt}
f - f_{ref} = \Delta f
\label{eq:freq_dev}
\end{equation}
\begin{equation}
\setlength{\abovedisplayskip}{3pt}
\setlength{\belowdisplayskip}{3pt}
P_{t} = (f - f_{ref})(Js + D) + P_{ref}.
\label{eq:inv_power}
\end{equation}

{\color{black}
Based on the derivations in \eqref{eq:freq_response} and \eqref{eq:freq_dev}, the dynamic frequency–power relation in \eqref{eq:inv_power} follows. The frequency used in the control equations is obtained from the phase-locked loop (PLL), and it hardly exerts destabilizing effects to the system under study. This  control method enables flexible power regulation by the inverter. When the grid frequency drops, the controller actively reduces power output to the HHESS, thereby alleviating stress on the grid and contributing to frequency support.}

In summary, the control strategy of HHESS begins by sampling the three-phase grid voltages and currents, which are then transformed into the $dq$-axis components through coordinate transformation. The inverter, operating under the inertia emulation control framework, generates a reference power signal based on the measured frequency deviation. Subsequently, a current control loop within the grid-connected converter regulates the output through PWM signal generation. This coordinated control process enables the proposed system to provide a rapid and effective response to grid frequency fluctuations.

{\color{black}
\vspace{-0.5em}
\section{Modular Scalability and Expansion Strategy}
%To facilitate practical deployment across various scales and application scenarios, the HHESS must offer modular scalability. Rather than relying on centralized coordination or extensive inter-device communication, the proposed control architecture supports plug-and-play expansion by treating each PEMEL, AEL, or SC module as functionally independent but behaviorally equivalent under the same control law.

Fig.~\ref{Fig. 7} presents the HHESS that incorporates $K$ PEMELs, $N$ AELs, and $M$ SCs. Leveraging the symmetry and parallel structure of these device groups, the entire system can be analytically reduced to a unified equivalent model comprising a single representative PEMEL, AEL, and SC unit, without loss of generality.

This structural simplification enables analytical derivation of the equivalent droop parameters for the expanded system as follows:

For the AEL group, by defining $\alpha_{\mathrm{eq}}$ as the equivalent droop coefficient, the group-level control parameter can be derived as:
    \begin{equation}
    \setlength{\abovedisplayskip}{3pt}
    \setlength{\belowdisplayskip}{3pt}
    \frac{1}{\alpha_{\mathrm{eq}}} = \sum_{n=1}^{N} \frac{1}{\alpha_n}.
    \label{eq:alpha_eq}
    \end{equation}

For the SC group, by defining $\zeta_{\mathrm{eq}}$ and $k_{\mathrm{eq}}$ as the equivalent CID control parameters, the group-level expression is derived as:
    \begin{equation}
    \setlength{\abovedisplayskip}{3pt}
    \setlength{\belowdisplayskip}{3pt}
    \begin{aligned}
    \frac{1}{s \zeta_{\mathrm{eq}}} \cdot \frac{s + k_{\mathrm{eq}}}{s} 
    &= \frac{1}{\sum_{m=1}^{M} s \zeta_m \cdot \frac{s}{s + k_m}} \\
    \Rightarrow \quad \frac{\zeta_{\mathrm{eq}}}{s + k_{\mathrm{eq}}} 
    &= \sum_{m=1}^{M} \frac{\zeta_m}{s + k_m}.
    \end{aligned} 
    \label{eq:zeta_eq}
    \end{equation}
    
\begin{figure}[t]
	\setlength{\abovecaptionskip}{-3pt}
	\setlength{\belowcaptionskip}{-5pt}
	\centering
	\includegraphics[scale=1]{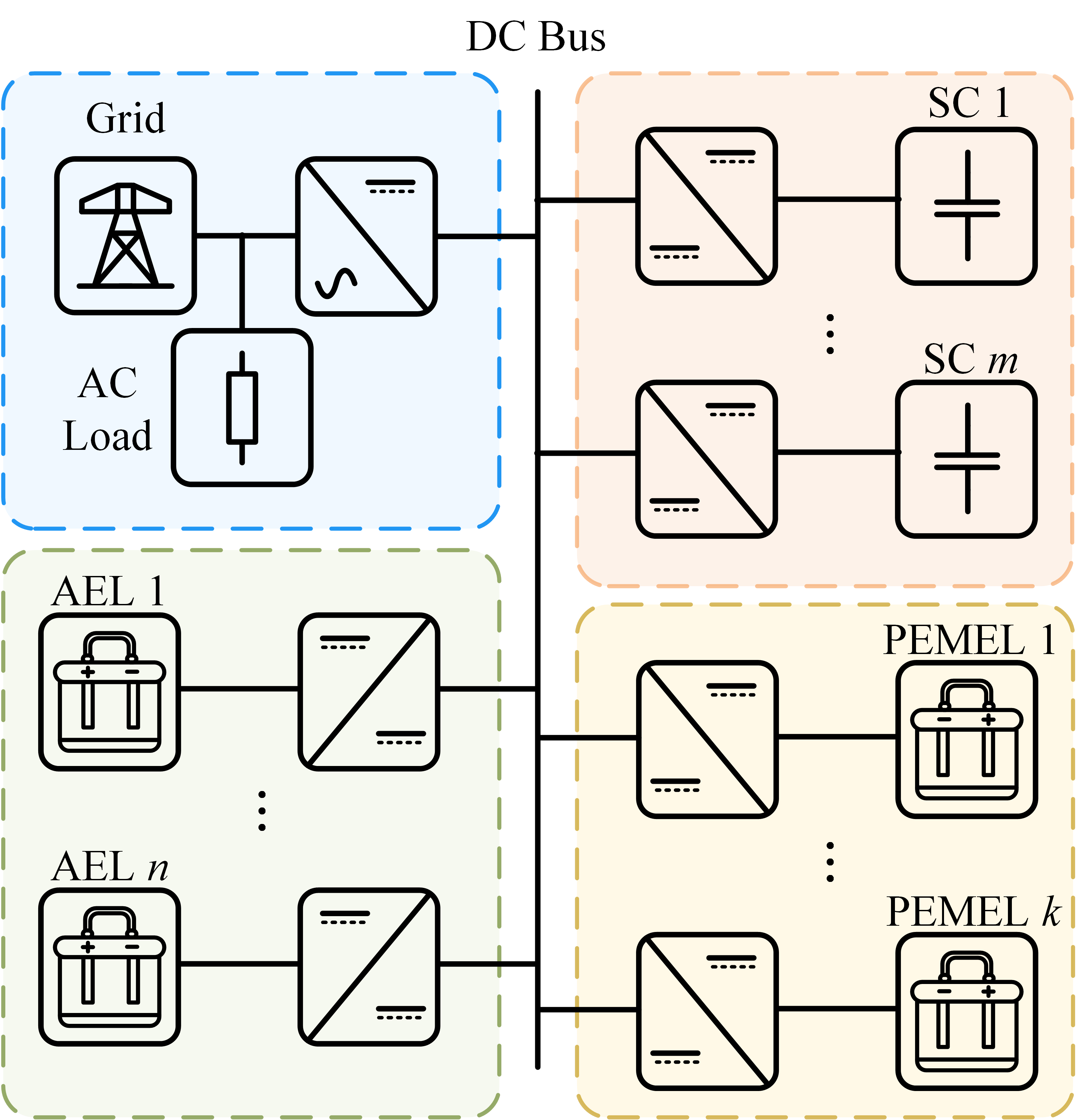}
	\caption{\textcolor{black}{HHESS Comprising $K$ PEMELs, $N$ AELs, and $M$ SCs.}}
	\label{Fig. 7}
	\vspace{-1em}
\end{figure}

If all SC branches use identical gain \( k_m = k \), then the equivalent parameters simplify to:
    \begin{equation}
    \setlength{\abovedisplayskip}{3pt}
    \setlength{\belowdisplayskip}{3pt}
    \zeta_{\mathrm{eq}} = \sum_{m=1}^{M} \zeta_m, \quad k_{\mathrm{eq}} = k.
    \label{eq:zeta_sum}
    \end{equation}

For the PEMEL group, by defining $\beta_{\mathrm{eq}}$ and $\gamma_{\mathrm{eq}}$ as the equivalent DID droop parameters, the group-level control expression is derived as:
    \begin{equation}
    \setlength{\abovedisplayskip}{3pt}
    \setlength{\belowdisplayskip}{3pt}
    \begin{aligned}
    \frac{1}{\frac{1}{\beta_{\mathrm{eq}}} + s \gamma_{\mathrm{eq}}} 
    &= \frac{1}{\sum_{k=1}^{K} \left( \frac{1}{\beta_k} + s \gamma_k \right)} \\
    \Rightarrow \quad \frac{1}{\beta_{\mathrm{eq}}} + s \gamma_{\mathrm{eq}} 
    &= \sum_{k=1}^{K} \left( \frac{1}{\beta_k} + s \gamma_k \right).
    \end{aligned}
    \label{eq:PEMEL_eq_impedance}
    \end{equation}
    
Therefore, the equivalent PEMEL parameters are calculated as:
    \begin{equation}
    \setlength{\abovedisplayskip}{3pt}
    \setlength{\belowdisplayskip}{3pt}
    \frac{1}{\beta_{\mathrm{eq}}} = \sum_{k=1}^{K} \frac{1}{\beta_k}, \quad
    \gamma_{\mathrm{eq}} = \sum_{k=1}^{K} \gamma_k.
    \label{eq:PEMEL_eq_params}
    \end{equation}

Whenever a new PEMEL unit $(K+1)$, AEL unit $(N+1)$, and SC unit $(M+1)$ are integrated into the system, the equivalent parameters of the HHESS must be updated accordingly. To preserve the dynamic performance of the original system, the control parameters of the newly added PEMEL, AEL, and SC modules must be appropriately recalibrated.

As previously discussed in Section~\ref{hhesscontrol}, define the practical power-sharing indices as $k_1 = \alpha / \beta$ and $k_2 = \zeta / \gamma$. $k_1$ denotes the steady-state power sharing ratio between the AEL and PEMEL groups and $k_2$ represents the transient power sharing ratio between the SC and PEMEL groups during frequency regulation. In order to maintain consistent dynamic behavior during system scaling, the values of $k_1$ and $k_2$ must remain constant throughout expansion.

Based on \eqref{eq:omega_xi_k1k2}, assuming uniform controller gain $k$ and constant ratios $k_1$, $k_2$, the overall system's natural frequency and damping ratio can be computed as:
\begin{equation}
\setlength{\abovedisplayskip}{3pt}
\setlength{\belowdisplayskip}{3pt}
{\omega _{0eq}} = \sqrt {\frac{{k{\kern 1pt} ({k_1} + 1)}}{{{\kern 1pt} (1 + {k_2}){\gamma _{eq}}{\alpha _{eq}}}}}
\label{eq:omega0}
\end{equation}
\begin{equation}
\setlength{\abovedisplayskip}{3pt}
\setlength{\belowdisplayskip}{3pt}
{\xi _{eq}} = \frac{1}{{2{\omega _{0eq}}}}\left( {\frac{k}{{{k_2} + 1}} + \frac{{{k_1} + 1}}{{({k_2} + 1){\gamma_{eq}}{\alpha _{eq}}}}} \right).
\label{eq:damping_ratio}
\end{equation}

To preserve identical dynamic characteristics before and after system expansion, it is necessary to ensure that both the natural frequency $\omega_{0eq}$ and damping ratio $\xi_{eq}$ remain unchanged. The following condition can be derived:
\begin{equation}
\setlength{\abovedisplayskip}{3pt}
\setlength{\belowdisplayskip}{3pt}
\gamma_{\mathrm{eq}} \, \alpha_{\mathrm{eq}} = \gamma_{\mathrm{eq'}} \, \alpha_{\mathrm{eq'}}
\label{eq:cp_rp_eq}.
\end{equation}

In the expanded system, the control coefficient $\alpha_{N+1}$ for the newly added PEMEL unit can be selected according to~\eqref{eq:AEL_alpha}. The updated equivalent droop parameters of the system can then be computed as follows:
\begin{equation}
\setlength{\abovedisplayskip}{3pt}
\setlength{\belowdisplayskip}{3pt}
\frac{1}{\alpha_{\mathrm{eq'}}} = \frac{1}{\alpha_{\mathrm{eq}}} + \frac{1}{\alpha_{N+1}}
\label{eq:rp_parallel}
\end{equation}
\begin{equation}
\setlength{\abovedisplayskip}{3pt}
\setlength{\belowdisplayskip}{3pt}
\gamma_{\mathrm{eq'}} = \gamma_{\mathrm{eq}} + \gamma_{K+1}.
\label{eq:cp_sum}
\end{equation}

Based on \eqref{eq:cp_rp_eq}, \eqref{eq:rp_parallel}, and \eqref{eq:cp_sum}, the required droop coefficient $\gamma_{K+1}$ for the newly added PEMEL unit is computed as:
\begin{equation}
\setlength{\abovedisplayskip}{3pt}
\setlength{\belowdisplayskip}{3pt}
\gamma_{K+1} = \frac{\alpha_{\mathrm{eq}}}{\alpha_{N+1}} \gamma_{\mathrm{eq}}.
\label{eq:cp_new}
\end{equation}

This parameter generalization ensures that dynamic performance and power allocation logic are preserved across system scales, offering a flexible and low-cost pathway to extend HHESS capacity with minimal modification to the existing control infrastructure.
}

\vspace{-1em}
\section{Large-Signal Stability Analysis With MPT}
Small-signal linearization is commonly employed to analyze system stability and facilitate parameter design in power systems\cite{Kot2019}. However, this approach is effective only for small perturbations around equilibrium points, making it insufficient for accurately predicting system stability under significant disturbances, such as voltage dips, load fluctuations, or system faults. Considering the HHESS studied in this paper is inherently nonlinear, large-signal stability analysis based on Lyapunov functions is essential to capture system behavior under major disturbances.
\vspace{-1em}
\subsection{Fundamentals of MPT}

The fundamental concept of large-signal stability analysis via Lyapunov functions is to construct an energy function suitable for the studied system. The large-signal stability criteria are derived by analyzing the positive definiteness of the Lyapunov energy function and the negative definiteness of its derivative. Mixed potential theory, first introduced by R.K. Brayton and J.K. Moser, is essentially a specialized form of the Lyapunov function\cite{RK1964}. Assuming the system contains $r$ inductive branches and $s$ capacitive branches, the mixed potential function $P(i,v)$ can be expressed as follows:
\begin{equation}
\setlength{\abovedisplayskip}{3pt}
\setlength{\belowdisplayskip}{3pt}
P(i,v) = \int \sum_{\mu > r + s} v_\mu \, d i_\mu + \sum_{r + s} i_\sigma v_\sigma.
\label{eq:power_expression}
\end{equation}

The mixed potential function $P(i,v)$ comprises two parts. In \eqref{eq:power_expression}, the term $\int \sum_{\mu>r+s} v_\mu di_\mu$ denotes the summation of the current potentials associated with all non-energy-storage components (such as resistors and non-ideal power sources) in the system, while the term $\sum_{\sigma=r+1}^{r+s} i_\sigma v_\sigma$ represents the total stored energy in all capacitive elements.

Since the mixed potential function $P(i,v)$ is constructed based on the potential functions of nonlinear circuit components such as resistors, inductors, capacitors, and other elements, it must satisfy the following relationship with the nonlinear circuit model to enable stability analysis using mixed potential theory:
\begin{equation}
\setlength{\abovedisplayskip}{3pt}
\setlength{\belowdisplayskip}{3pt}
\left\{
\begin{aligned}
L_\rho \frac{d i_\rho}{dt} &= \frac{\partial P(i,v)}{\partial i_\rho} \\
C_\sigma \frac{d v_\sigma}{dt} &= \frac{\partial P(i,v)}{\partial v_\sigma}.
\end{aligned}
\right.
\label{eq:dynamics_iv}
\end{equation}

In \eqref{eq:dynamics_iv}, $i_\rho$ denotes the currents in inductive branches ($\rho=1,\dots,r$), and $v_\sigma$ represents the voltages in capacitive branches ($\sigma=r+1,\dots,r+s$). \eqref{eq:dynamics_iv} must hold true for the mixed potential theory to be applicable. Thus, the general form of the mixed potential function $P(i,v)$ can be expressed as:
\begin{equation}
\setlength{\abovedisplayskip}{3pt}
\setlength{\belowdisplayskip}{3pt}
P(i,v) = -A(i) + B(v) + (i, Dv).
\label{eq:power_func}
\end{equation}

In \eqref{eq:power_func}, $A(i)$ denotes the current potential function of non-energy-storage elements, $B(v)$ represents the voltage potential function of non-energy-storage elements, and the term $(i,Dv)$ represents the energy associated with capacitive elements and some non-energy-storage components in the circuit. Using the mixed potential stability criterion, a function is constructed as shown in \eqref{eq:P_star}:
\begin{equation}
\setlength{\abovedisplayskip}{3pt}
\setlength{\belowdisplayskip}{3pt}
\begin{split}
P^{*}(i,v) =\; & \frac{\mu_{1}+\mu_{2}}{2}\,P(i,v)\;+\;
\frac{1}{2}\,(P_i,\;L^{-1}P_i)\; \\
& +\; \frac{1}{2}\,(P_v,\;C^{-1}P_v)\;\longrightarrow\;\infty.
\end{split}
\label{eq:P_star}
\end{equation}

The partial derivatives in \eqref{eq:P_star} are defined as 
\( P_i = \partial P(i,v)/\partial i \), 
\( P_v = \partial P(i,v)/\partial v \), 
\( A_{ii}(i) = \partial^2 A/\partial i^2 \), 
and \( B_{vv}(v) = \partial^2 B/\partial v^2 \). 
\( \mu_1 \) denote the smallest eigenvalue of the matrix 
\( L^{-1/2} A_{ii}(i) L^{-1/2} \), 
and \( \mu_2 \) denote the smallest eigenvalue of the matrix 
\( C^{-1/2} B_{vv}(v) C^{-1/2} \).

When the condition $\mu_1 + \mu_2 > 0$ is satisfied, it guarantees that all trajectories of the nonlinear system eventually converge to the steady-state operating point as $|i| + |v| \rightarrow \infty$, thereby ensuring system stability. 

% \vspace{-1em}
\subsection{Large-Signal Stability Analysis for HHESS}
 \begin{figure}[t]
	\setlength{\abovecaptionskip}{-3pt}
	\setlength{\belowcaptionskip}{-5pt}
	\centering
	\includegraphics[scale=1]{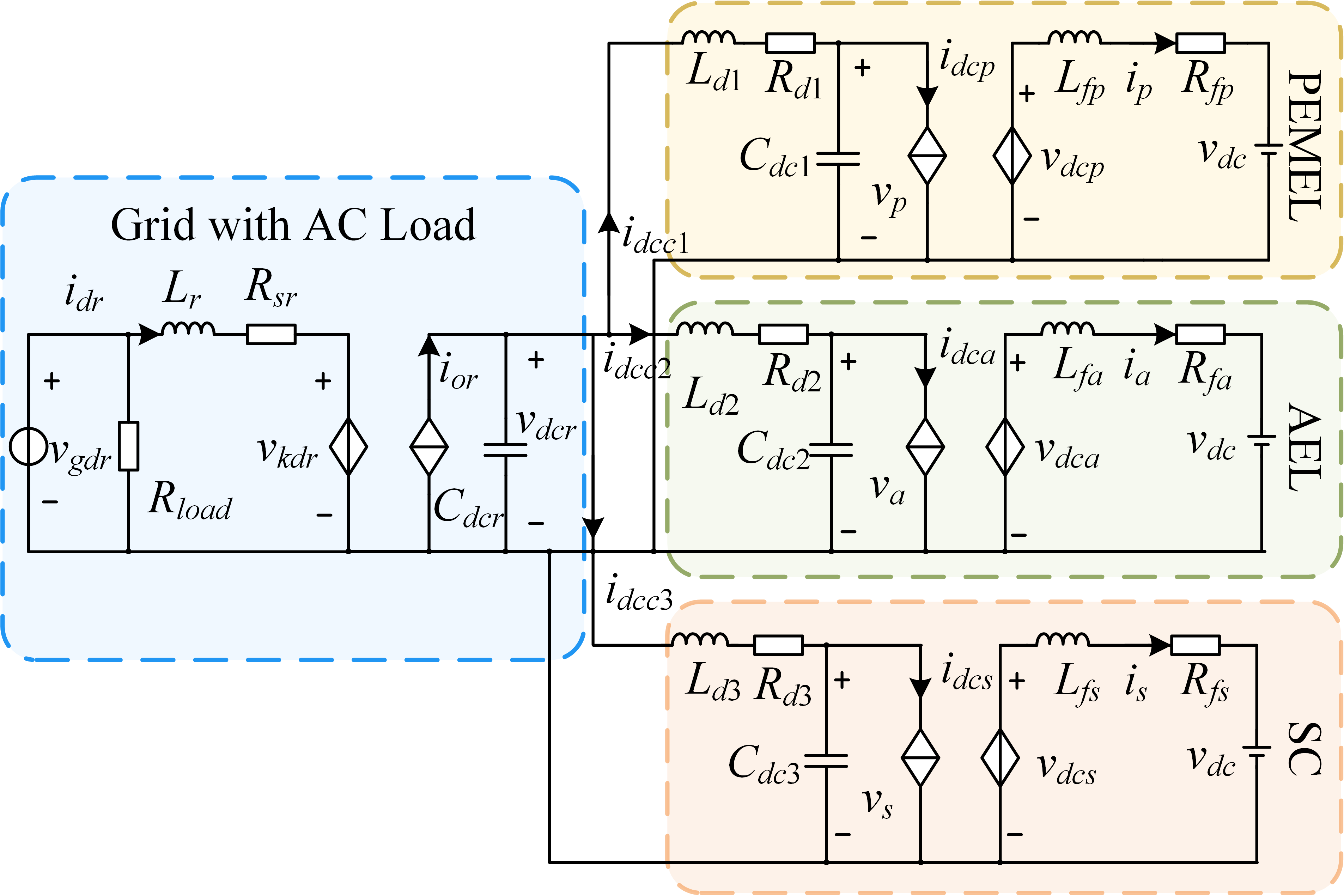}
	\caption{Simplified average value model of HHESS.}
	\label{Fig. 9}
	\vspace{-1em}
\end{figure}

For the large-signal stability analysis of the HHESS, a simplified average model is established as shown in Fig.~\ref{Fig. 9}. Detailed derivations of the model can be found in \cite{HuajunZheng2020}. Upon Fig.~\ref{Fig. 9}, the averaged model of the HHESS is analyzed to construct the mixed potential function \( P(i, v) \). Specifically, the summation of current potential functions for all non-energy-storage components, as well as the total energy stored in all capacitive elements, is first calculated:
\begin{comment}
\begin{equation}
\setlength{\abovedisplayskip}{3pt}
\setlength{\belowdisplayskip}{3pt}
\begin{aligned}
\int \sum_{\mu} v_\mu \, d i_\mu =\;
& \int_0^{i_{dr}} (v_{gdr} - v_{kdr} - R_{sr} i_{dr}) \, d i_{dr}
+ \int_0^{i_{or}} v_{dcr} \, d i_{or} \\
& - \int_0^{i_{dcc1}} R_{d1} i_{dcc1} \, d i_{dcc1}
- \int_0^{i_{dcp}} v_{dc1} \, d i_{dcp} \\
& - \int_0^{i_{dcc2}} R_{d2} i_{dcc2} \, d i_{dcc2}
- \int_0^{i_{dca}} v_{dc2} \, d i_{dca} \\
& - \int_0^{i_{dcc3}} R_{d3} i_{dcc3} \, d i_{dcc3}
- \int_0^{i_{dcs}} v_{dc3} \, d i_{dcs} \\
& + \int_0^{i_p} (v_{dcp} - v_{dc} - R_{fp} i_p) \, d i_p \\
& + \int_0^{i_a} (v_{dca} - v_{dc} - R_{fa} i_a) \, d i_a \\
& + \int_0^{i_s} (v_{dcs} - v_{dc} - R_{fs} i_s) \, d i_s
\end{aligned}
\label{eq:power_integral_two_per_line}
\end{equation}  
\end{comment}

\begin{equation}
\setlength{\abovedisplayskip}{3pt}
\setlength{\belowdisplayskip}{3pt}
\begin{aligned}
\int \sum_{\mu} v_\mu \, d i_\mu =\;
& \int_0^{i_{dr}} (v_{gdr} - v_{kdr} - R_{sr} i_{dr}) \, d i_{dr} \\
&- \sum_{j=1}^{3} \int_0^{i_{dccj}} R_{dj} i_{dccj} \, d i_{dccj} \\
&+ \sum_{k = p,a,s} \int_0^{i_k} (v_{dck} - v_{dc} - R_{fk} i_k) \, d i_k \\
&+ \int_0^{i_{or}} v_{dcr} \, d i_{or} - \sum_{\substack{k = p,a,s}} \int_0^{i_{dck}} v_{k} \, d i_{dck}
\end{aligned}
\label{eq:power_integral_two_per_line}
\end{equation}

\begin{comment}
\begin{equation}
\setlength{\abovedisplayskip}{3pt}
\setlength{\belowdisplayskip}{3pt}
\begin{aligned}
\sum_{\sigma = r + 1}^{r + s} v_\sigma i_\sigma =\; 
& v_{dcr} (i_{dcc1} + i_{dcc2} + i_{dcc3} - i_{or}) \\
& + v_{dc1} (i_{dcp} - i_{dcc1}) 
+ v_{dc2} (i_{dca} - i_{dcc2}) \\
& + v_{dc3} (i_{dcs} - i_{dcc3})
\end{aligned}
\label{eq:vi_sum}
\end{equation}    
\end{comment}

\begin{equation}
\setlength{\abovedisplayskip}{3pt}
\setlength{\belowdisplayskip}{3pt}
\begin{aligned}
\sum_{\sigma = r + 1}^{r + s} v_\sigma i_\sigma =\;
& v_{dcr} \left( \sum_{j=1}^{3} i_{dccj} - i_{or} \right) \\
& {}+ \sum_{\substack{k = p,a,s}} v_{k} \left(i_{dck} - i_{dccj}\right).
\end{aligned}
\label{eq:vi_sum}
\end{equation}

After simplification, the mixed potential function \( P(i, v) \) is obtained. Next, based on \eqref{eq:dynamics_iv}, the system is verified to satisfy the conditions required by the mixed potential theory, leading to \eqref{eq:partial_derivatives1} and \eqref{eq:partial_derivatives2}, which confirms the validity of the approach.

\begin{comment}

\begin{equation}
\setlength{\abovedisplayskip}{3pt}
\setlength{\belowdisplayskip}{3pt}
\left\{
\begin{aligned}
\frac{\partial P(i,v)}{\partial i_{dr}}     &= v_{gdr} - v_{kdr} - R_{sr} \cdot i_{dr} = L_r \frac{d i_{dr}}{dt} \\
\frac{\partial P(i,v)}{\partial i_{dcc1}}   &= - R_{d1} \cdot i_{dcc1} + v_{dcr} - v_{dc1} = L_{d1} \frac{d i_{dcc1}}{dt} \\
\frac{\partial P(i,v)}{\partial i_{dcc2}}   &= - R_{d2} \cdot i_{dcc2} + v_{dcr} - v_{dc2} = L_{d2} \frac{d i_{dcc2}}{dt} \\
\frac{\partial P(i,v)}{\partial i_{dcc3}}   &= - R_{d3} \cdot i_{dcc3} + v_{dcr} - v_{dc3} = L_{d3} \frac{d i_{dcc3}}{dt} \\
\frac{\partial P(i,v)}{\partial i_p}        &= v_{dcp} - v_{dc} - R_{fp} \cdot i_p = L_{fp} \frac{d i_p}{dt} \\
\frac{\partial P(i,v)}{\partial i_a}        &= v_{dca} - v_{dc} - R_{fa} \cdot i_a = L_{fa} \frac{d i_a}{dt} \\
\frac{\partial P(i,v)}{\partial i_s}        &= v_{dcs} - v_{dc} - R_{fs} \cdot i_s = L_{fs} \frac{d i_s}{dt} \\
\frac{\partial P(i,v)}{\partial v_{dcr}}    &= - i_{or} + i_{dcc1} + i_{dcc2} + i_{dcc3} = - C_{dcr} \frac{d v_{dcr}}{dt} \\
\frac{\partial P(i,v)}{\partial v_{dc1}}    &= i_{dcp} - i_{dcc1} = - C_{dc1} \frac{d v_{dc1}}{dt} \\
\frac{\partial P(i,v)}{\partial v_{dc2}}    &= i_{dca} - i_{dcc2} = - C_{dc2} \frac{d v_{dc2}}{dt} \\
\frac{\partial P(i,v)}{\partial v_{dc3}}    &= i_{dcs} - i_{dcc3} = - C_{dc3} \frac{d v_{dc3}}{dt}
\end{aligned}
\right.
\label{eq:partial_derivatives}
\end{equation}
\end{comment}

\begin{equation}
\setlength{\abovedisplayskip}{3pt}
\setlength{\belowdisplayskip}{3pt}
\left\{
\begin{aligned}
\frac{\partial P(i,v)}{\partial i_{dr}}     &= v_{gdr} - v_{kdr} - R_{sr} \cdot i_{dr} = L_r \frac{d i_{dr}}{dt} \\
\frac{\partial P(i,v)}{\partial i_{dccj}}   &= - R_{dj} \cdot i_{dccj} + v_{dcr} - v_{k} = L_{dj} \frac{d i_{dccj}}{dt} \\
\frac{\partial P(i,v)}{\partial i_k}        &= v_{dck} - v_{dc} - R_{fk} \cdot i_k = L_{fk} \frac{d i_k}{dt} \\
 & j = 1,2,3;\quad k = p,a,s
\end{aligned}
\right.
\label{eq:partial_derivatives1}
\end{equation}

\begin{equation}
\setlength{\abovedisplayskip}{3pt}
\setlength{\belowdisplayskip}{3pt}
\left\{
\begin{aligned}
\frac{\partial P(i,v)}{\partial v_{dcr}} &= - i_{or} + \sum_{j=1}^{3} i_{dccj} = - C_{dcr} \frac{d v_{dcr}}{dt} \\
\frac{\partial P(i,v)}{\partial v_{k}} &= i_{dck} - i_{dccj} = - C_{k} \frac{d v_{k}}{dt} \\
& j = 1,2,3;\quad k = p,a,s
\end{aligned}
\right.
\label{eq:partial_derivatives2}
\end{equation}

 Accordingly, the mixed potential function of the system is reformulated into the form of \eqref{eq:power_func}, from which the expressions of $A(i)$ and $B(v)$ can be obtained, as shown in \eqref{eq:A_current} and \eqref{eq:B_voltage}.
\begin{comment}
\begin{equation}
\setlength{\abovedisplayskip}{3pt}
\setlength{\belowdisplayskip}{3pt}
\begin{aligned}
A(i) =\;& 
- \int_0^{i_{dr}} \left( v_{gdr} - v_{kdr} \right) d i_{dr}
+ \frac{1}{2} R_{sr} i_{dr}^2 \\
&+ \frac{1}{2} R_{d1} i_{dcc1}^2 
+ \frac{1}{2} R_{d2} i_{dcc2}^2 
+ \frac{1}{2} R_{d3} i_{dcc3}^2 \\
&- \int_0^{i_p} (v_{dcp} - v_{dc})\, d i_p 
+ \frac{1}{2} R_{fp} i_p^2 \\
&- \int_0^{i_a} (v_{dca} - v_{dc})\, d i_a 
+ \frac{1}{2} R_{fa} i_a^2 \\
&- \int_0^{i_s} (v_{dcs} - v_{dc})\, d i_s 
+ \frac{1}{2} R_{fs} i_s^2
\end{aligned}
\label{eq:A_current}
\end{equation}    
\end{comment}

\begin{equation}
\setlength{\abovedisplayskip}{3pt}
\setlength{\belowdisplayskip}{3pt}
\begin{aligned}
A(i) =\;& 
- \int_0^{i_{dr}} \left( v_{gdr} - v_{kdr} \right) d i_{dr}
+ \frac{1}{2} R_{sr} i_{dr}^2 \\[4pt]
&+ \sum_{j=1}^{3} \frac{1}{2} R_{dj} i_{dccj}^2 
+ \sum_{k=p,a,s} \frac{1}{2} R_{fk} i_k^2\\[4pt]
&- \sum_{k=p,a,s} \int_0^{i_k} (v_{dck} - v_{dc}) \, d i_k 
\end{aligned}
\label{eq:A_current}
\end{equation}

\begin{comment}
\begin{equation}
\setlength{\abovedisplayskip}{3pt}
\setlength{\belowdisplayskip}{3pt}
\begin{aligned}
B(v) =\;
& - \int_0^{v_{dcr}} i_{or}\, d v_{dcr}
+ \int_0^{v_{dc1}} i_{dcp}\, d v_{dc1} \\
& + \int_0^{v_{dc2}} i_{dca}\, d v_{dc2}
+ \int_0^{v_{dc3}} i_{dcs}\, d v_{dc3}
\end{aligned}
\label{eq:B_voltage}
\end{equation}    
\end{comment}

\begin{equation}
\setlength{\abovedisplayskip}{3pt}
\setlength{\belowdisplayskip}{3pt}
\begin{aligned}
B(v) =\;
& - \int_0^{v_{dcr}} i_{or} \, d v_{dcr}
+ \sum_{\substack{k = p,a,s}} \int_0^{v_{k}} i_{dck} \, d v_k.
\end{aligned}
\label{eq:B_voltage}
\end{equation}

By taking the second-order partial derivatives of \eqref{eq:A_current} and\eqref{eq:B_voltage}, the matrix expressions \( A_{ii}(i) \) and \( B_{vv}(v) \) are obtained, corresponding respectively to the system’s inductance matrix and capacitance matrix.
\begin{equation}
\begin{aligned}
A_{ii}(i) =\;
\operatorname{diag}\!\Bigl(
& \frac{\partial v_{kdr}}{\partial i_{dr}} + R_{sr},\;
R_{d1},\;
R_{d2},\;
R_{d3},\;
-\frac{\partial v_{dcp}}{\partial i_{p}}\\[2pt]
& + R_{fp}, -\frac{\partial v_{dca}}{\partial i_{a}} + R_{fa},\;
-\frac{\partial v_{dcs}}{\partial i_{s}} + R_{fs}
\Bigr)
\end{aligned}
\label{eq:Aii_diag}
\end{equation}
\begin{equation}
B_{vv}(v) = \operatorname{diag}\!\Bigl(
- \tfrac{\partial i_{or}}{\partial v_{dcr}},\;
\tfrac{\partial i_{dcp}}{\partial v_{p}},\;
\tfrac{\partial i_{dca}}{\partial v_{a}},\;
\tfrac{\partial i_{dcs}}{\partial v_{s}}
\Bigr)
\label{eq:Bvv_diag}
\end{equation}
\begin{equation}
L = \operatorname{diag}\!\bigl(
L_{r},\;
L_{d1},\;
L_{d2},\;
L_{d3},\;
L_{fp},\;
L_{fa},\;
L_{fs}
\bigr)
\label{eq:L_diag}
\end{equation}
\begin{equation}
C = \operatorname{diag}\!\bigl(
C_{dcr},\;
C_{dc1},\;
C_{dc2},\;
C_{dc3}
\bigr).
\label{eq:C_diag}
\end{equation}

Then, combining these results with the control strategies introduced in Section~\ref{sec:control}, the smallest eigenvalues of the matrices $L^{-1/2} A_{ii}(i) L^{-1/2}$ and $C^{-1/2} B_{vv}(v) C^{-1/2}$ can be calculated as follows (equations \eqref{eq:mu1_definition}–\eqref{eq:X3}):
\begin{equation}
\begin{aligned}
\mu_1 = \min\Biggl(
& \frac{k_{ipr} \left( \dfrac{D (V_{gdr} - I_{drref} R_{sr}) }{V_{gdr}D_{grid}} + 1 \right)}{L_r \left( 1 - k_{ipr} \dfrac{D I_{drref} }{V_{gdr}D_{grid}} \right)} 
+ \frac{R_{sr}}{L_r}, \\[3pt]
& \frac{R_{d1}}{L_{d1}},\;
\frac{R_{d2}}{L_{d2}},\;
\frac{R_{d3}}{L_{d3}},\;
\frac{k_{ip1} V_{dc1} + R_{fp}}{L_{fp}}, \\[3pt]
& \frac{k_{ip2} V_{dc2} + R_{fa}}{L_{fa}},\;
\frac{k_{ip3} V_{dc3} + R_{fs}}{L_{fs}}
\Biggr)
\end{aligned}
\label{eq:mu1_definition}
\end{equation}
\begin{equation}
\begin{aligned}
\mu_2 = \min\Biggl(
& \frac{3(V_{gdr} - I_{drref} R_{sr}) I_{drref}}{2C_{dcr} V_{dcr}^2},\;
\frac{X_1}{C_{dc1}},  \frac{X_2}{C_{dc2}},\;
\frac{X_3}{C_{dc3}}
\Biggr).
\end{aligned}
\label{eq:mu2_definition}
\end{equation}

{\color{black}
In \eqref{eq:mu1_definition}–\eqref{eq:mu2_definition}, $k_{ipr}$ denotes the proportional gain of the current loop in the AC/DC control, while $k_{ip1}$, $k_{ip2}$, and $k_{ip3}$ are the proportional gains of the inner current loops of the PEMEL, AEL, and SC DC/DC stages, respectively. The coefficient $D$ is the virtual damping gain introduced in Section~III. $D_{\mathrm{grid}}$ represents the damping characteristic of the low-inertia grid. $I_{drref}$ is the grid-side $d$-axis current reference and $V_{gdr}$ is the $d$-axis grid voltage. All necessary parameters have been provided in Table~\ref{tab:hhess_params}. The auxiliary terms $X_1$, $X_2$, and $X_3$ are compact combinations of system parameters. Their explicit definitions are as follows:
}
\begin{equation}
\begin{aligned}
X_1 =&\; \frac{d_p - k_{ip1} I_p}{(R_{fp} + k_{ip1} V_{p})^2} \cdot \bigl( - R_{fp} + k_{ip1} I_{\text{pref}} R_{fp} \\
&- k_{ip1} V_{dc} + k_{ip1} V_{p} + k_{ip1}^2 V_{p} I_{\text{pref}} \bigr)
\end{aligned}
\label{eq:X1}
\end{equation}
\begin{equation}
\begin{aligned}
X_2 =&\; \frac{d_a - k_{ip2} I_a}{(R_{fa} + k_{ip2} V_{a})^2} \cdot \bigl( - R_{fa} + k_{ip2} I_{\text{aref}} R_{fa} \\
&- k_{ip2} V_{dc} + k_{ip2} V_{a} + k_{ip2}^2 V_{a} I_{\text{aref}} \bigr)
\end{aligned}
\label{eq:X2}
\end{equation}
\begin{equation}
\begin{aligned}
X_3 =&\; \frac{d_s - k_{ip3} I_s}{(R_{fs} + k_{ip3} V_{s})^2} \cdot \bigl( - R_{fs} + k_{ip3} I_{\text{sref}} R_{fs} \\
&- k_{ip3} V_{dc} + k_{ip3} V_{s} + k_{ip3}^2 V_{s} I_{\text{sref}} \bigr).
\end{aligned}
\label{eq:X3}
\end{equation}

\begin{figure}[t]
	\setlength{\abovecaptionskip}{-3pt}
	\setlength{\belowcaptionskip}{-3pt}
	\centering
	\includegraphics[scale=1]{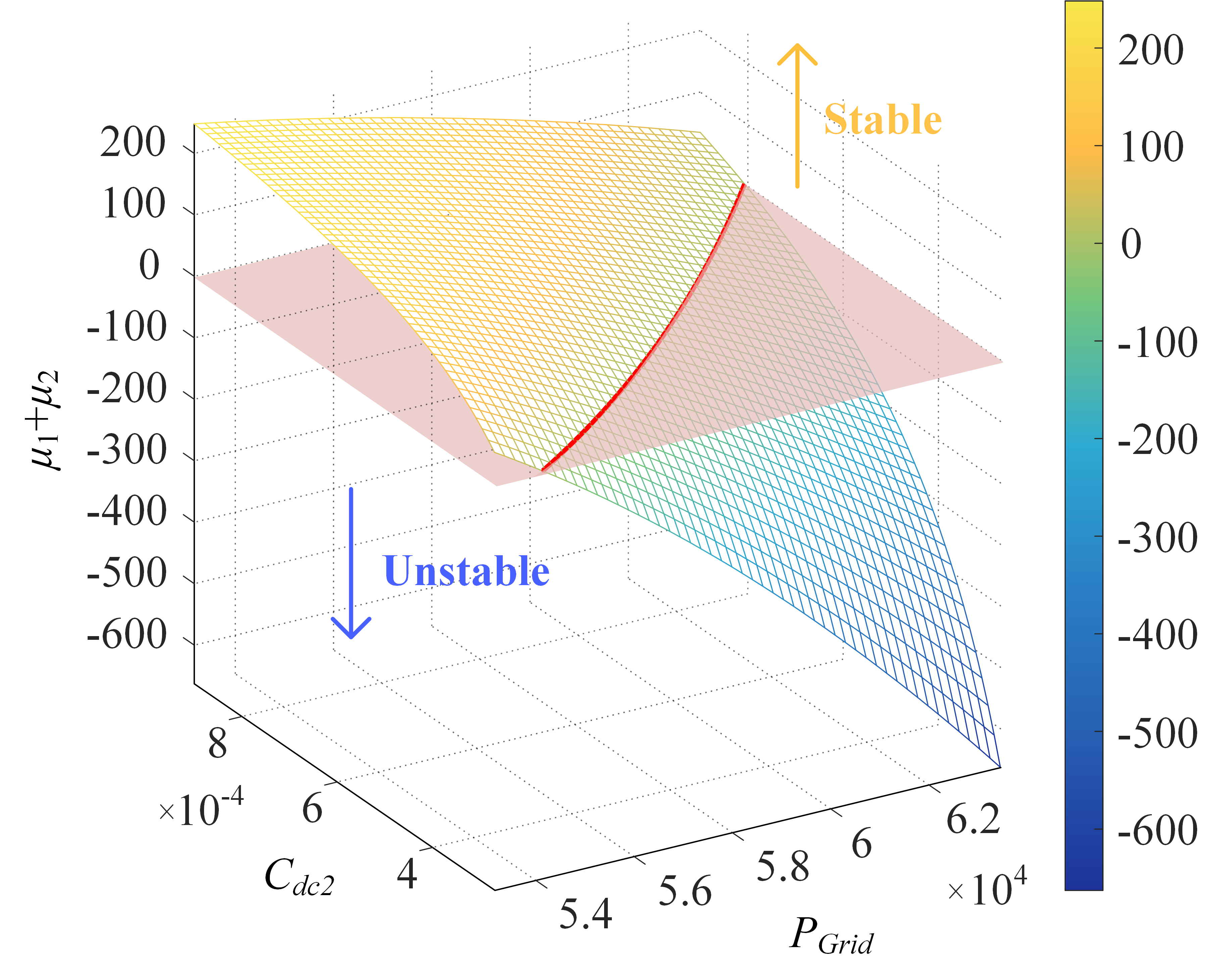}
	\caption{Stability analysis of HHESS with $P_{\mathrm{Grid}}$ and $C_{\mathrm{dc2}}$.}
	\label{Fig. 10}
	\vspace{-1em}
\end{figure}

{\color{black}
Having derived the smallest eigenvalues of these two matrices, the large-signal stability criterion $\mu_1+\mu_2>0$ is obtained. Based on this criterion, the HHESS stability region can be identified. 

Taking the total grid power $P_{\mathrm{Grid}}$ and the DC-side capacitance $C_{dc2}$ as an example, Fig.~\ref{Fig. 10} shows the corresponding stability boundary. To generate this figure, $\mu_1$+$\mu_2$ is computed from \eqref{eq:mu1_definition} and \eqref{eq:mu2_definition} over the specified parameter ranges, with $P_{\mathrm{Grid}}$ varied from 54kW to 62kW and $C_{dc2}$ from 200$\mu$F to 1000$\mu$F, while keeping all other parameters unchanged. The locus $\mu_1$+$\mu_2$=0 (red curve) separates stable and unstable regions; increasing $C_{dc2}$ or reducing $P_{\mathrm{Grid}}$ moves the operating point deeper into the stable region, whereas simultaneously high $P_{\mathrm{Grid}}$ and small $C_{dc2}$ narrow the margin. Thus, the MPT-based criterion provides an efficient way to delineate admissible parameter ranges, facilitating stability assessment and guiding controller/plant tuning under large disturbances.
}

\vspace{-0.5em}
\section{Experiments}
\begin{figure}[t]
	\setlength{\abovecaptionskip}{-3pt}
	\setlength{\belowcaptionskip}{-5pt}
	\centering
	\includegraphics[scale=1]{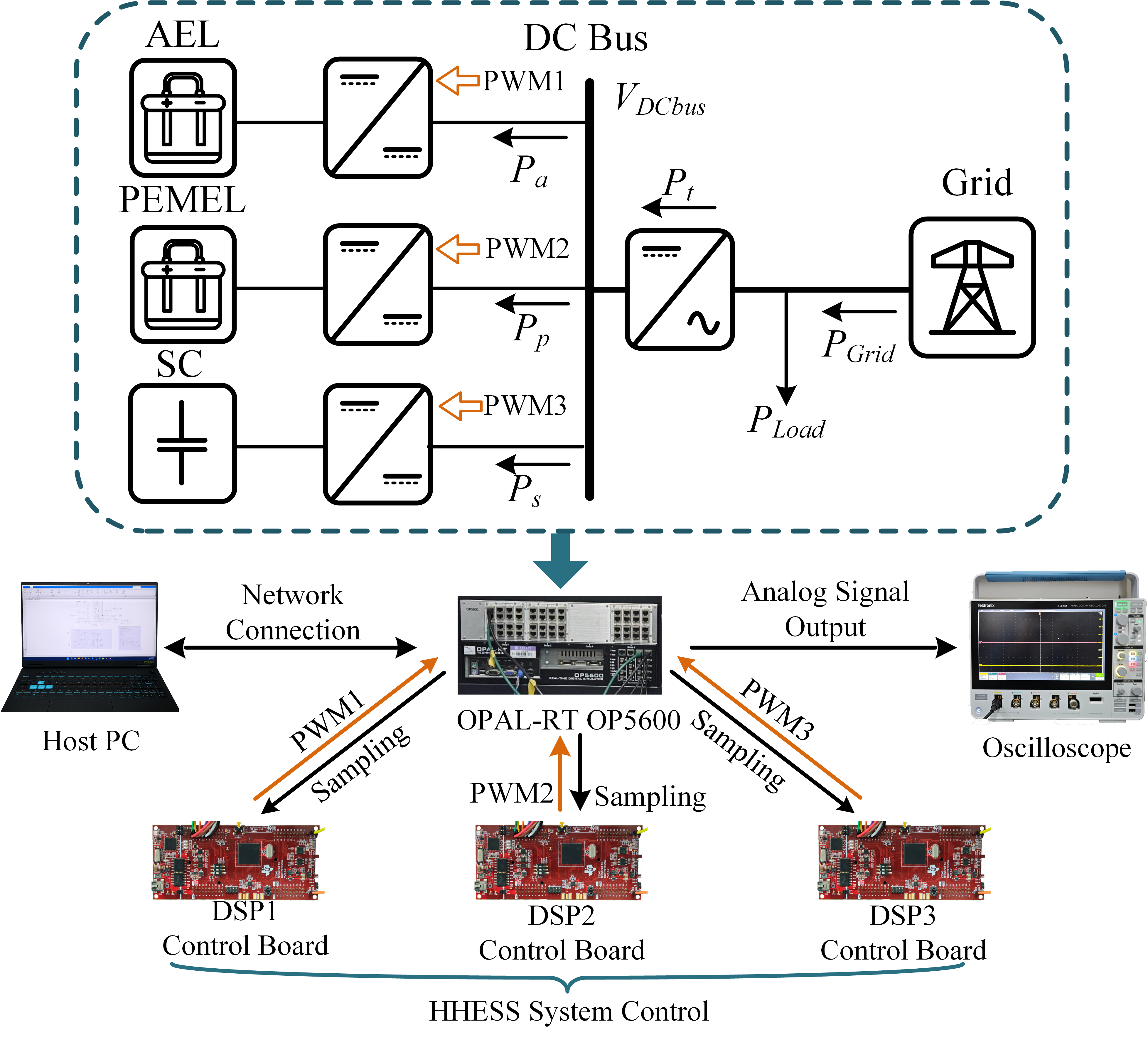}
	\caption{In-house hardware-in-the-loop (HIL) experimental platform.}
	\label{Fig. 11}
	\vspace{0em}
\end{figure}

{\color{black}
To corroborate the system functions of the HHESS regulated by the proposed control strategy, as well as to verify the effectiveness and reliability of the large-signal stability analysis in determining critical parameter boundaries, a hardware-in-the-loop (HIL) platform is established on an OPAL-RT OP5600, as shown in Fig.~\ref{Fig. 11}. In the HIL model, the PEMEL, AEL, and SC branches are represented as DC sources to streamline testing while preserving the system-level response characteristics. All other key components, including the three bidirectional DC/DC converters, the grid-tied DC/AC inverter and grid interface, the DC‑bus capacitor, and the line impedances, are modeled in real time and are not omitted. Specifically, AEL, PEMEL, and SC are interfaced with the common DC bus through three independent DC/DC converters and then connected to the grid via a grid-tied inverter. The analog input/output range of the OP5600 is ±16V; therefore, all physical quantities observed in the experiments are numerically scaled to conform to this range. All control parameters used in the experiments follow Table~\ref{tab:hhess_params}.}
\begin{table}[htbp]
    \vspace{0em}
    
    \caption{\color{black} Relevant System Parameters of HHESS}
    \label{tab:hhess_params}
    \setlength{\abovecaptionskip}{-5pt}
    \setlength{\belowcaptionskip}{-10pt}
    \vspace{-1em}
    \centering
    \setlength{\tabcolsep}{5pt} % 缩小列间距（默认约 6pt）
    \small  
    {\color{black}%% 略微缩小字号
    \begin{adjustbox}{max width=\linewidth}
    \begin{tabular}{c l c}
        \toprule  
        Parameter & Description & Value \\
        \midrule
        $V_{g\mathrm{r}}$ & Grid RMS voltage & 220 V \\
        $V_{\mathrm{ref}}$ & DC bus reference voltage & 750 V \\
        $C_{\mathrm{dcr}}$ & Inverter-side capacitor & 470 $\mu$F \\
        $C_{\mathrm{dci}} \ i=1,2,3$ & DC-side capacitor voltages & 470 $\mu$F \\
        $\alpha$ & Droop parameter of AEL & 6.67e-4 V/W\\
        $\beta, \gamma$ & Droop parameters of PEMEL & 6.67e-4 V/W ,\;750 W·s/V\\
        $\zeta,\ k$ & Droop parameters of SC & 1500 W·s/V,\;1 s$^{-1}$ \\
        $J,\ D$ & Inertia and damping factor & 100 W·s/Hz,\;3000 W/Hz \\
        $k_{ip1}, k_{ii1}$ & PI gain of DC/DC current loop for PEMEL & 0.0445,\;205.1 \\
        $k_{vp1}, k_{vi1}$ & PI gain of DC/DC voltage loop for PEMEL & 1.307,\;401.6 \\
        $k_{ip2}, k_{ii2}$ & PI gain of DC/DC current loop for AEL & 0.0178,\;32.8 \\
        $k_{vp2}, k_{vi2}$ & PI gain of DC/DC voltage loop for AEL & 0.392,\;36.1 \\
        $k_{ip3}, k_{ii3}$ & PI gain of DC/DC current loop for SC & 0.0445,\;205.1 \\
        $k_{vp3}, k_{vi3}$ & PI gain of DC/DC voltage loop for SC & 1.960,\;903.6 \\
        \bottomrule
    \end{tabular}
    \end{adjustbox}
    }% end color group
    \vspace{-1em}
\end{table}

\vspace{-1em}
\subsection{Case 1: HHESS with Step-Up Load Disturbance}
\begin{figure}[t]
	\setlength{\abovecaptionskip}{-3pt}
	\setlength{\belowcaptionskip}{-5pt}
	\centering
	\includegraphics[scale=1]{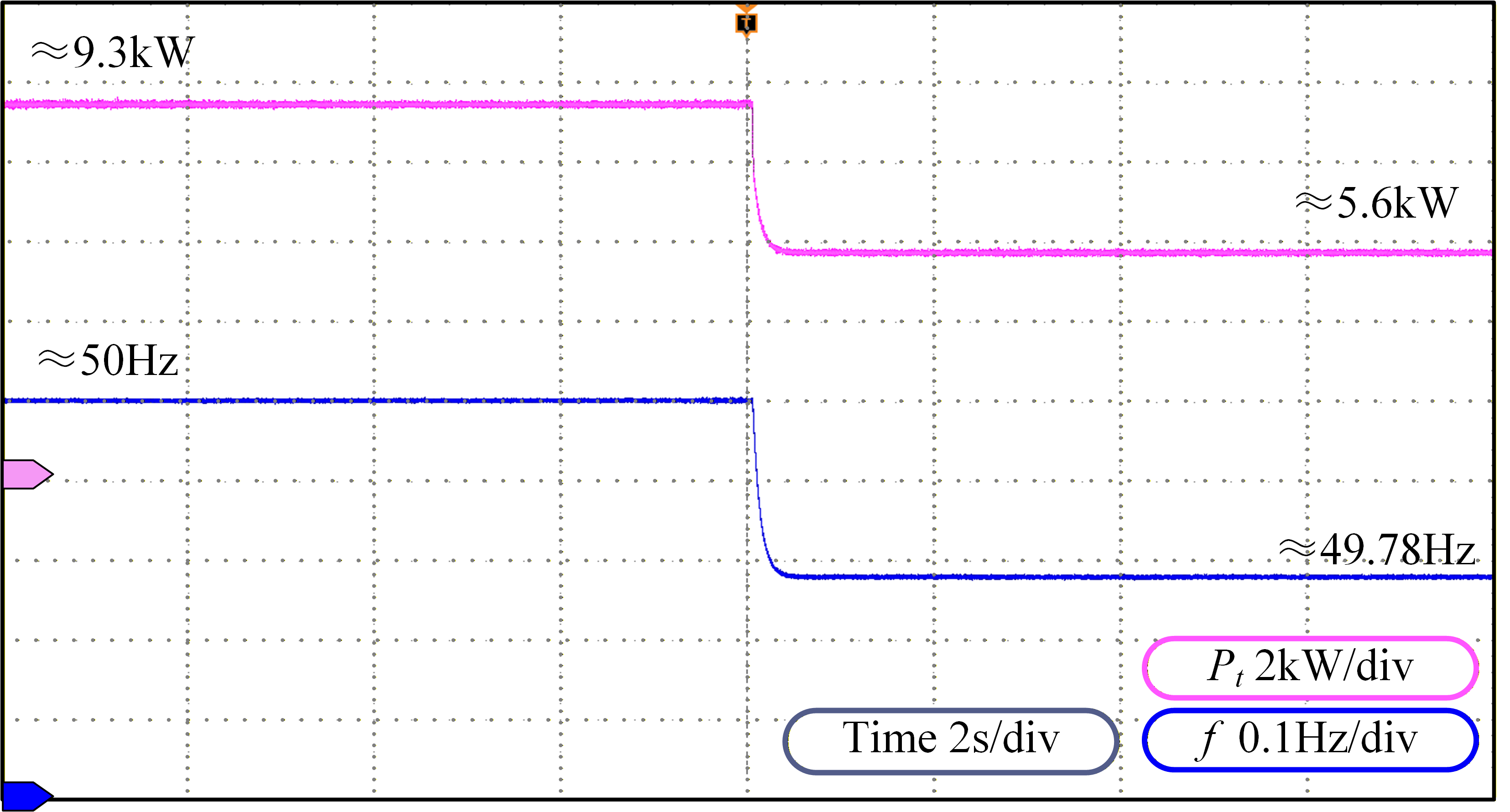}
	\caption{\textcolor{black}{ $P_{\text{t}}$ and $f$ under step-up load disturbance.}}
	\label{Fig. 13}
	\vspace{-1em}
\end{figure}

\begin{figure}[t]
	\setlength{\abovecaptionskip}{-3pt}
	\setlength{\belowcaptionskip}{-5pt}
	\centering
	\includegraphics[scale=1]{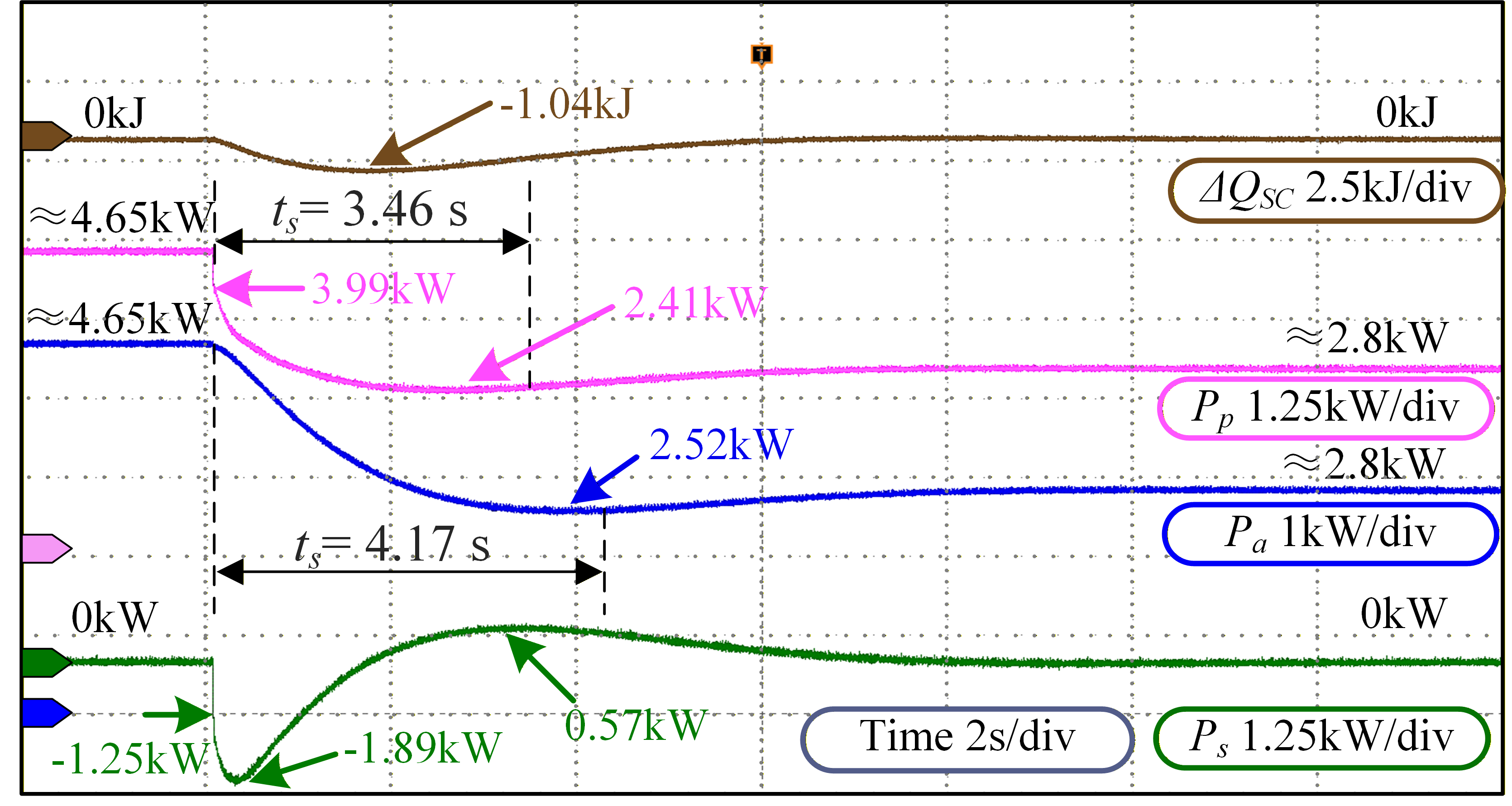}
	\caption{\textcolor{black}{Dynamic response of HHESS components under step-up load disturbance.}}
	\label{Fig. 12}
	\vspace{-1em}
\end{figure}

{\color{black}
 To investigate the dynamic performance of the HHESS under sudden increases in AC load power, a step disturbance is introduced by raising the load power $P_{\text{load}}$ from approximately 20kW to 33kW. As shown in Fig.~\ref{Fig. 13}, during the load step-up process, the dynamic responses of the total HHESS output power $P_{\text{t}}$ and the grid frequency $f$ are recorded. When the load increases abruptly, the inverter-controlled $P_{\text{t}}$ rapidly drops from approximately 9.3kW to 5.6kW. This fast reduction effectively relieves the burden on the grid and mitigates the frequency decline, resulting in a frequency drop from 50Hz to only 49.78Hz. 

Fig.~\ref{Fig. 12} further illustrates the internal power responses of individual components in the HHESS during the disturbance. The PEMEL reacts swiftly to the power variation, with $P_{\text{p}}$ falling from 4.65kW to roughly 2.8kW. During this transition, the power briefly dips to a minimum of 2.41kW, and the settling time is measured to be 3.46s. The calculated undershoot is approximately 13.9\%. In contrast, the AEL responds more slowly, yet its power $P_{\text{a}}$ eventually follows the same downward trend from 4.65kW to around 2.8kW. The lowest point during the fluctuation is 2.52kW, and the corresponding settling time is 4.17s. The calculated undershoot in this case is approximately 10\%. Simultaneously, the SC rapidly releases transient power to compensate for the sharp gap. During the fluctuation, its power reaches a minimum of -1.89kW and a peak of 0.57kW. As the system settles into a new steady state, the SC output gradually returns to zero.  The energy variation of the SC, denoted by $\Delta Q_{\text{SC}}$, is also shown in Fig.~\ref{Fig. 12}. It can be clearly observed that the SC’s net energy variation during the disturbance reaches a maximum discharge of 1.04kJ. However, the energy before and after the disturbance remains balanced, indicating that the proposed control strategy successfully achieves autonomous SOC recovery for the SC.
}

 \begin{figure}[!htbp]
	\setlength{\abovecaptionskip}{-3pt}
	\setlength{\belowcaptionskip}{-5pt}
	\centering
	\includegraphics[scale=1]{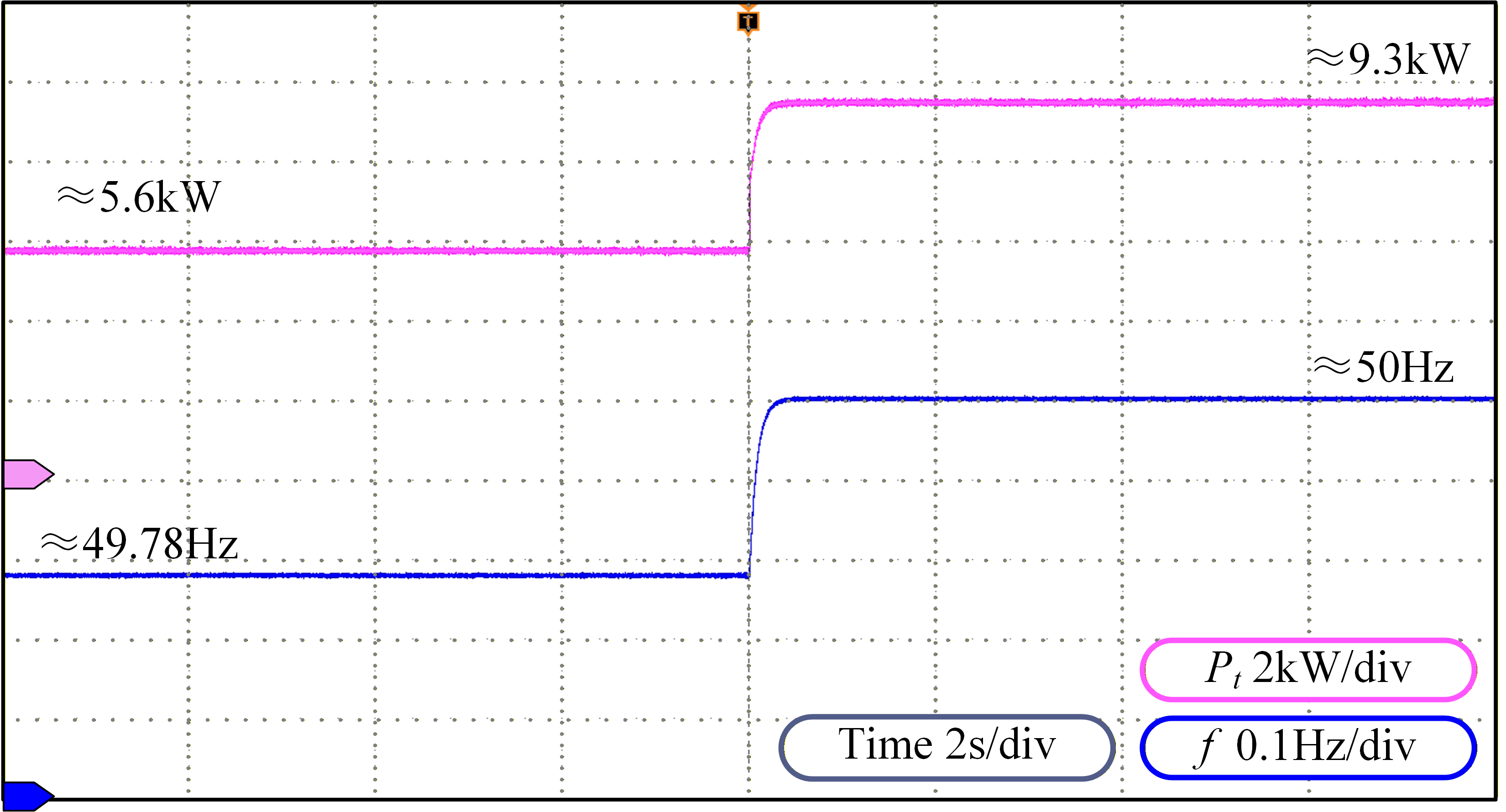}
	\caption{\textcolor{black}{$P_{\text{t}}$ and $f$ under step-down load disturbance.}}
	\label{Fig. 15}
	\vspace{-1em}
\end{figure}
\begin{figure}[!htbp]
	\setlength{\abovecaptionskip}{-3pt}
	\setlength{\belowcaptionskip}{-5pt}
	\centering
	\includegraphics[scale=1]{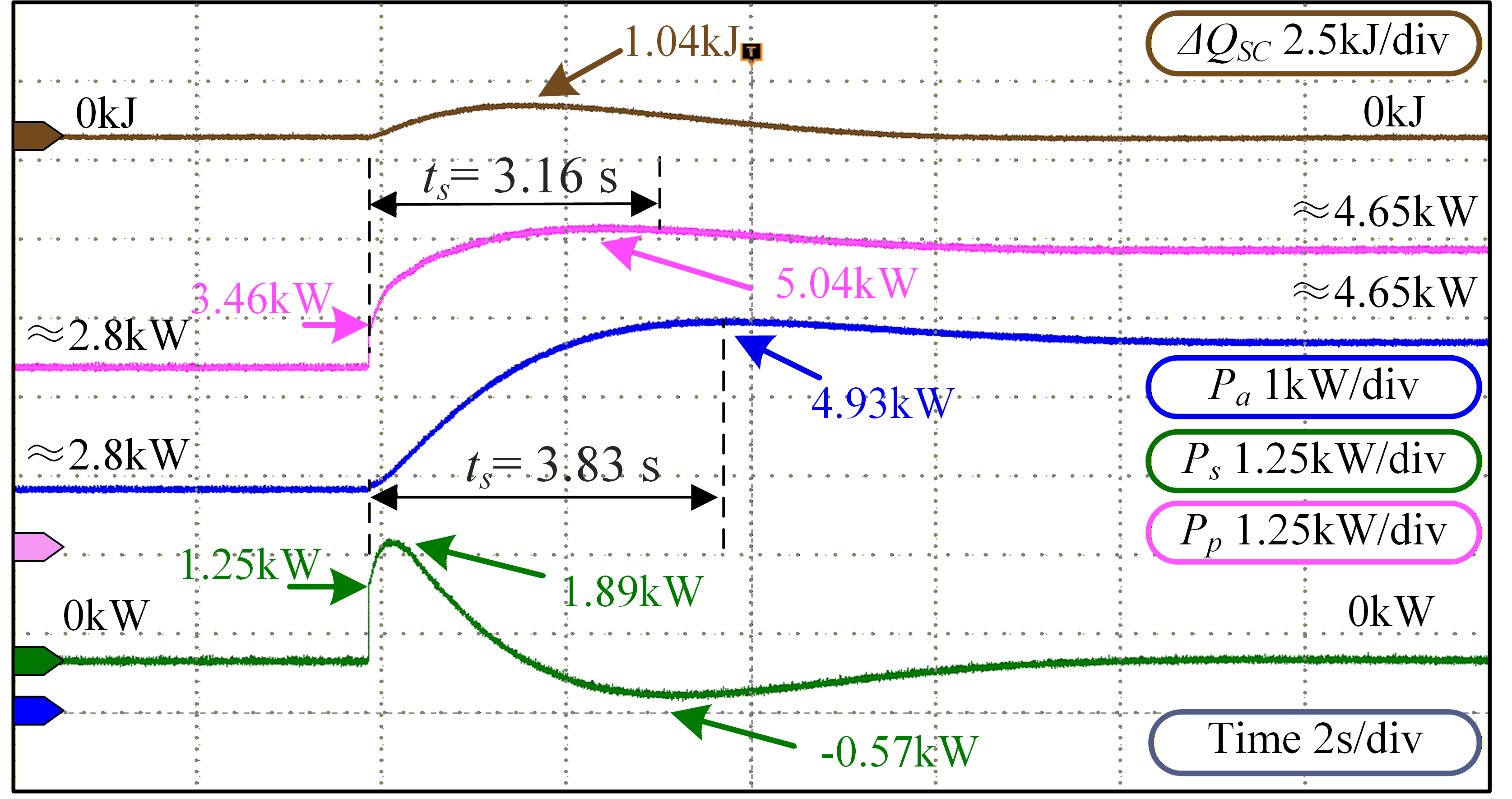}
	\caption{\textcolor{black}{Dynamic response of HHESS components under step-down load disturbance.}}
	\label{Fig. 14}
	\vspace{-1em}
\end{figure}
\begin{figure}[t]
	\setlength{\abovecaptionskip}{-3pt}
	\setlength{\belowcaptionskip}{-5pt}
	\centering
	\includegraphics[scale=1]{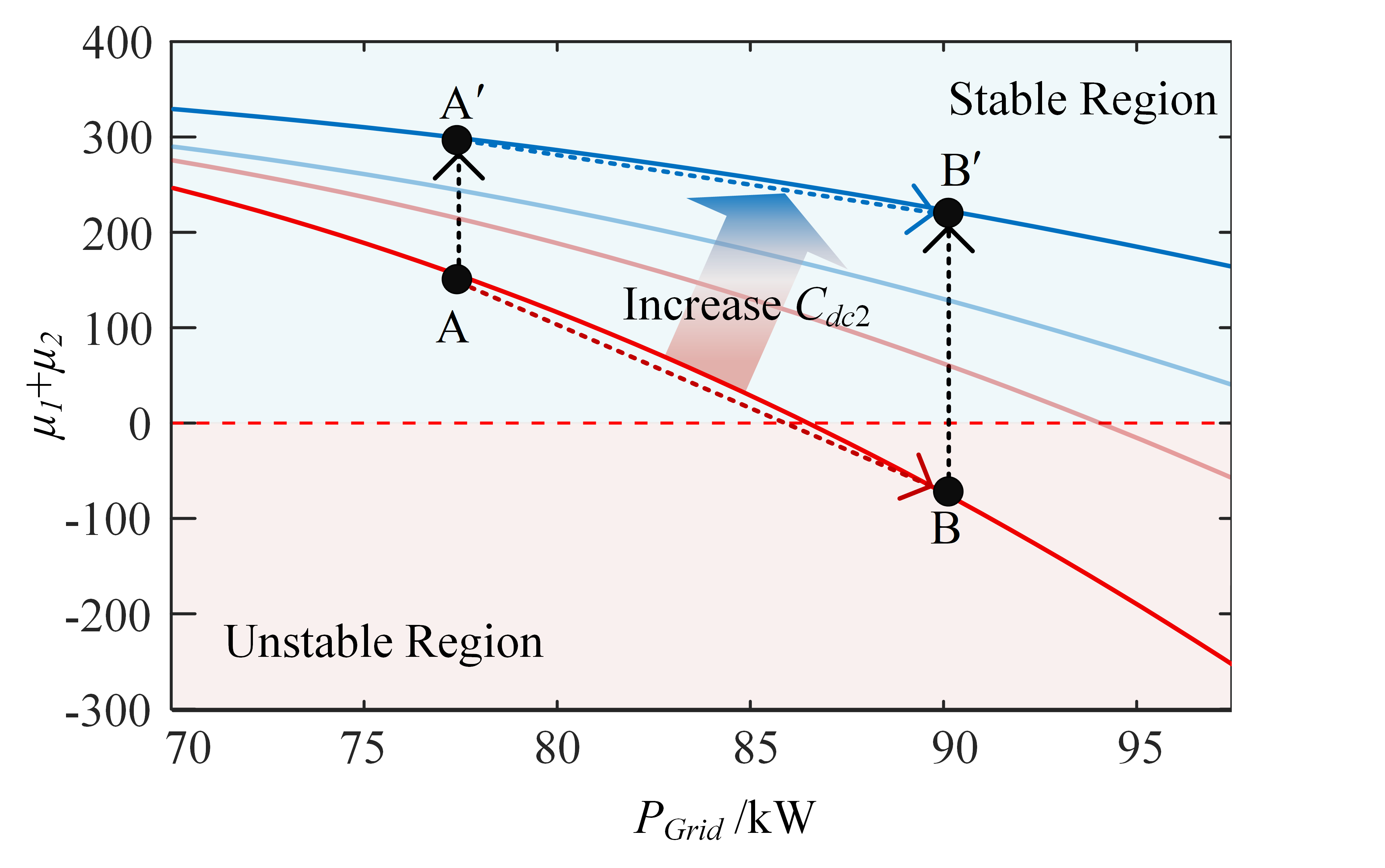}
	\caption{Analysis of large-signal stability boundary extension in HHESS based on $P_{\text{Grid}}$.}
	\label{Fig. 16}
	\vspace{-1em}
\end{figure}
\vspace{-1em}
\subsection{Case 2: HHESS with Step-Down Load Disturbance}
{\color{black}
Following Case 1, Case 2 examines the dynamic performance of the HHESS when the $P_{\text{load}}$ undergoes a sudden step-down disturbance from approximately 33kW back to around 20kW. As depicted in Fig.~\ref{Fig. 15}, when the load power abruptly decreases, the inverter-controlled HHESS power $P_{\text{t}}$ correspondingly recovers from about 5.6kW back up to approximately 9.3kW. Consequently, the grid frequency, initially stabilized at 49.78Hz, quickly returns to 50Hz after the disturbance. 

Fig.~\ref{Fig. 14} further illustrates the detailed power response of each HHESS component under this step-down load scenario. Similar to Case 1, the PEMEL rapidly reacts to the load change, with its power $P_{\text{p}}$ promptly increasing from around 2.8kW to approximately 4.65kW. During the transient, the power peaks at 5.04kW, and the settling time is measured to be 3.16s. The calculated overshoot is therefore approximately 8.4\%. Conversely, the AEL exhibits a gradual increase in power from approximately 2.8kW to about 4.65kW , with a peak value of 4.93kW during the fluctuation and a settling time of 3.83s. The calculated overshoot in this case is approximately 6\%. Meanwhile, the SC instantly absorbs transient excess power caused by the sudden load decrease. Its power reaches a peak of 1.89kW and a minimum of -0.57kW, then gradually returns to zero as the system settles into a new steady state. Likewise, the SC’s net energy variation during the disturbance reaches a maximum absorption of 1.04kJ. However, the energy levels before and after the disturbance remain balanced, confirming that the proposed control strategy enables autonomous SOC recovery for the SC.
}

\begin{figure}[t]
	\setlength{\abovecaptionskip}{-3pt}
	\setlength{\belowcaptionskip}{-5pt}
	\centering
	\includegraphics[scale=1]{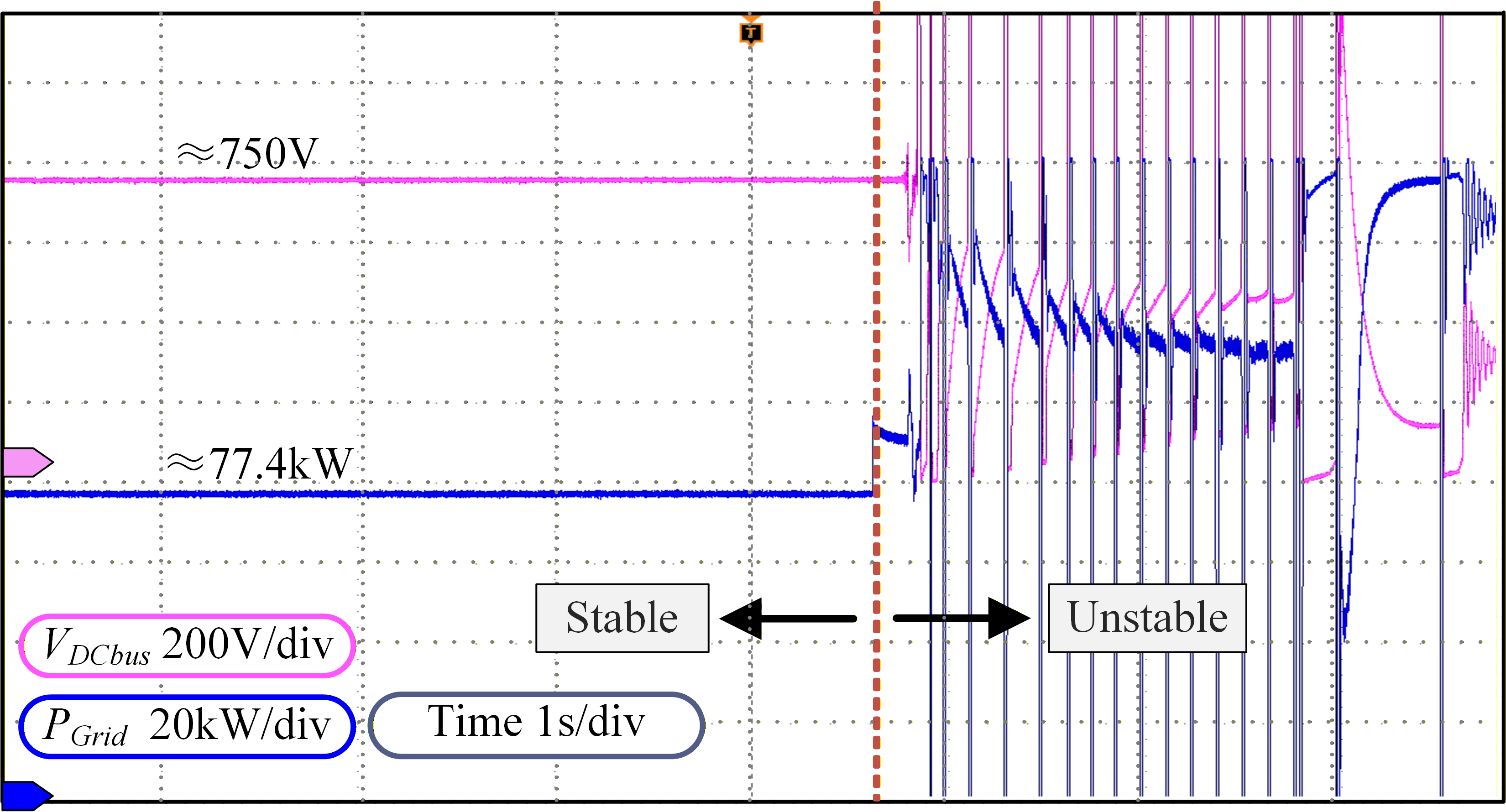}
	\caption{\textcolor{black}{$P_{\text{Grid}}$ and $V_{\text{DCbus}}$ when $C_{dc2}$=$470\,\mu\text{F}$.}}
	\label{Fig. 17}
	\vspace{-1em}
\end{figure}
\begin{figure}[t]
	\setlength{\abovecaptionskip}{-3pt}
	\setlength{\belowcaptionskip}{-5pt}
	\centering
	\includegraphics[scale=1]{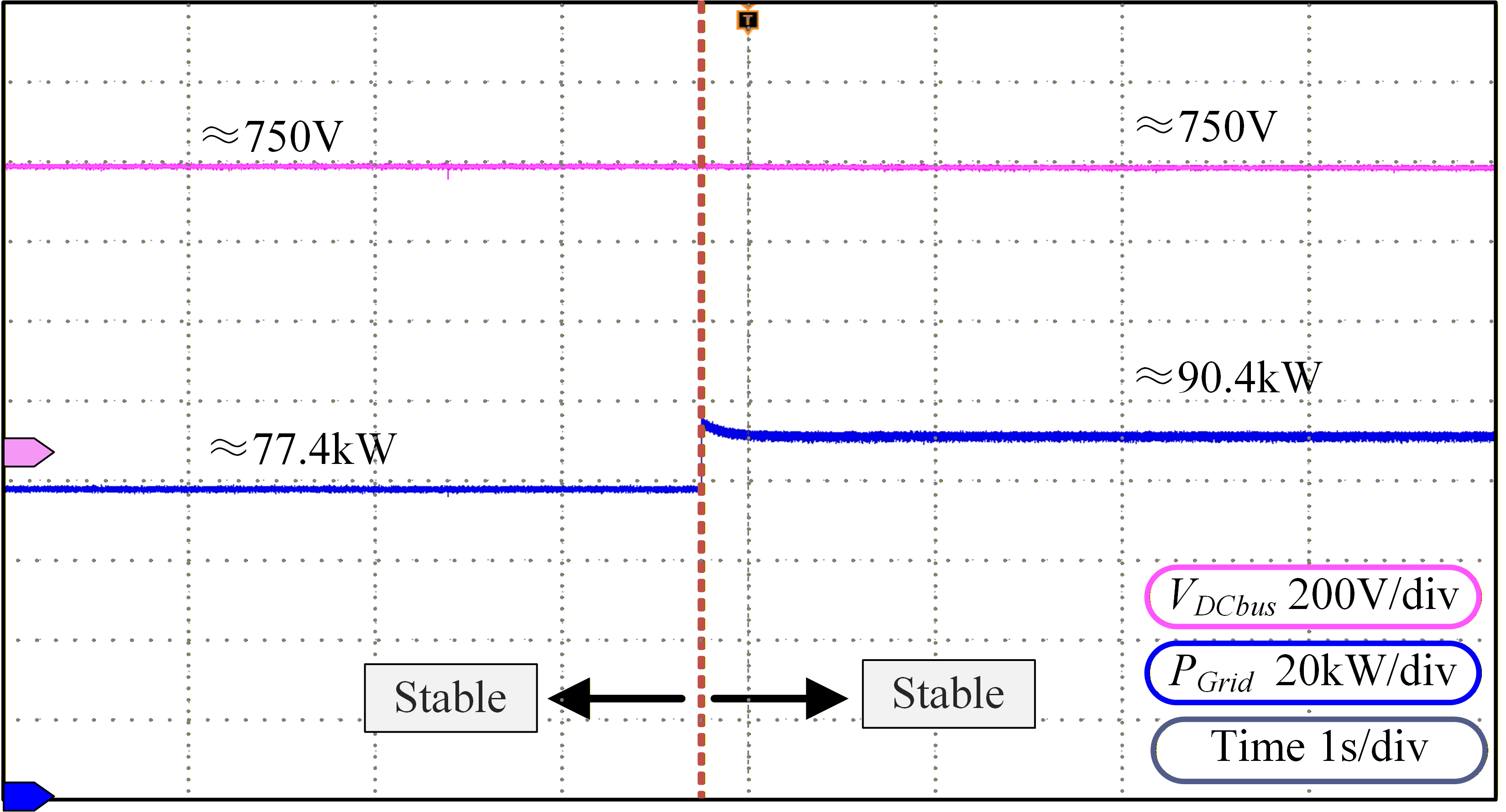}
	\caption{\textcolor{black}{$P_{\text{Grid}}$ and $V_{\text{DCbus}}$ when $C_{dc2}$=$4700\,\mu\text{F}$.}}
	\label{Fig. 18}
	\vspace{-1em}
\end{figure}

\vspace{-1em}
\subsection{Case 3: Verification of Large-Signal Stability Criterion}
Unlike the previous two cases, Case 3 aims to verify the accuracy of the large-signal stability criterion established in Section IV. To investigate the impact of grid total power ($P_{\text{Grid}}$) variations on system stability, a stability boundary diagram is constructed in Fig.~\ref{Fig. 16}. Initially, the operating point is set at 77.4 kW (Point A), which lies within the stable region. Following an abrupt load increase that raises the grid power to 90.4 kW, the operating state transitions from stable Point A into unstable Point B. By increasing the capacitance $C_{dc2}$ from $470\,\mu\text{F}$ to $4700\,\mu\text{F}$, the stability boundary shifts upward, moving the initial and final operating points from A to A' and from B to B', respectively. Under these new conditions, both points remain within the stable region, even after the load disturbance.

{\color{black}
Based on the stability analysis and predictive results shown in Fig.~\ref{Fig. 16}, experimental tests are carried out by introducing a disturbance in the AC load, which causes the $P_{\text{Grid}}$ to step up from 77.4kW to 90.4kW. Different capacitance values for $C_{dc2}$ are then applied to validate the system’s dynamic behavior. As illustrated in Fig.~\ref{Fig. 17}, with $C_{dc2}$ set at $470\,\mu\text{F}$, the initial $P_{\text{Grid}}$ and DC bus voltage $V_{\text{DCbus}}$ remain stable at approximately 77.4 kW and 750 V, respectively. However, once the $P_{\text{Load}}$ experiences a step increase, both $P_{\text{Grid}}$ and $V_{\text{DCbus}}$ rapidly become unstable, aligning with the analytical prediction in Fig.~\ref{Fig. 16}, where the system transitions from stable Point A to unstable Point B. In contrast, Fig.~\ref{Fig. 18} shows the response when $C_{dc2}$ is increased to $4700\,\mu\text{F}$. Under identical initial conditions, following the same load disturbance, the $P_{\text{Grid}}$ smoothly transitions from 77.4 kW to approximately 90.4 kW without instability. Meanwhile, the $V_{\text{DCbus}}$ remains steadily around 750 V, corroborating the analytical prediction from Fig.~\ref{Fig. 16} that increasing $C_{dc2}$ ensures continuous operation within the stable region, even under large-signal disturbances. In fact, $C_{dc2}$ can be further tuned so that, in the post‐disturbance regime, the stability index $\mu_{1}+\mu_{2}$ operates closer to the boundary. Such tuning preserves stability while enabling a tighter, more optimized selection of system parameters.
}

\begin{figure}[t]
	\setlength{\abovecaptionskip}{-3pt}
	\setlength{\belowcaptionskip}{-5pt}
	\centering
	\includegraphics[scale=0.9]{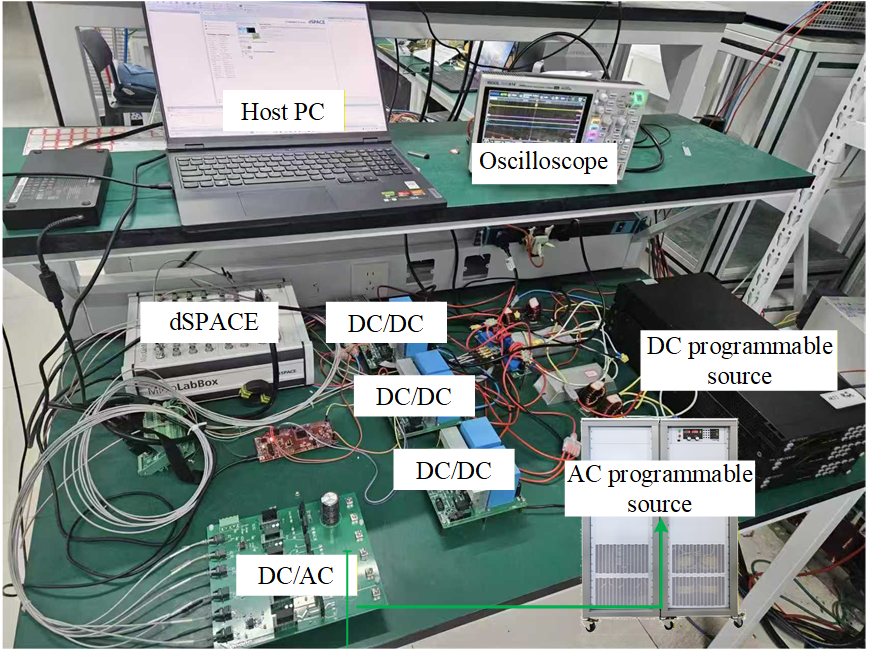}
	\caption{\textcolor{black}{Experimental platform.}}
	\label{Fig. 19}
	\vspace{-1em}
\end{figure}
\begin{figure}[t]
	\setlength{\abovecaptionskip}{-3pt}
	\setlength{\belowcaptionskip}{-5pt}
	\centering
	\includegraphics[scale=1]{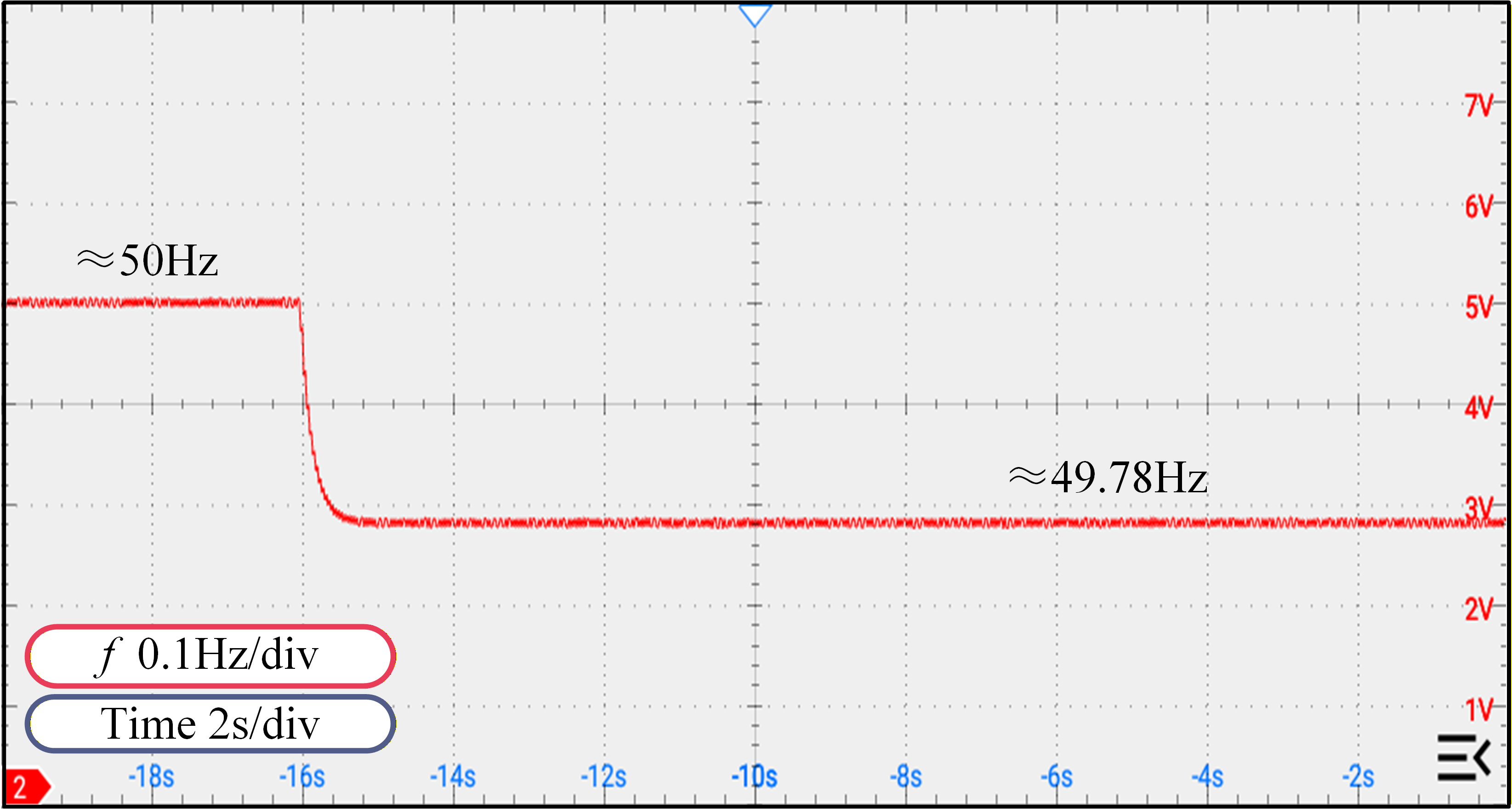}
	\caption{\textcolor{black}{Experimental results of grid frequency.}}
	\label{Fig. 20}
	\vspace{-1em}
\end{figure}
\begin{figure}[!htbp]
	\setlength{\abovecaptionskip}{-3pt}
	\setlength{\belowcaptionskip}{-5pt}
	\centering
	\includegraphics[scale=1]{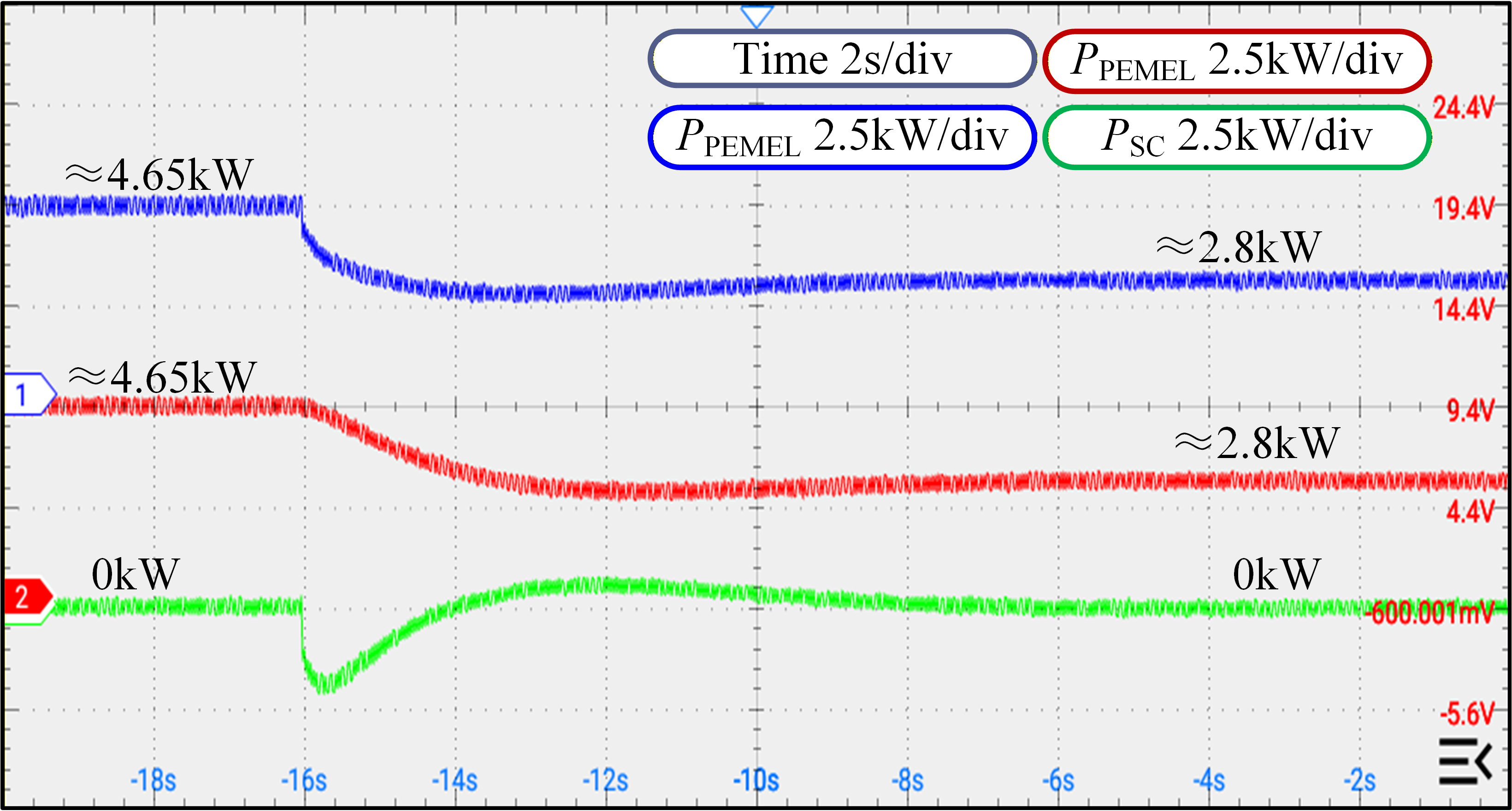}
	\caption{\textcolor{black}{Experimental results of HHESS components.}}
	\label{Fig. 21}
	\vspace{-1em}
\end{figure}
\begin{figure}[!htbp]
	\setlength{\abovecaptionskip}{-3pt}
	\setlength{\belowcaptionskip}{-5pt}
	\centering
	\includegraphics[scale=1]{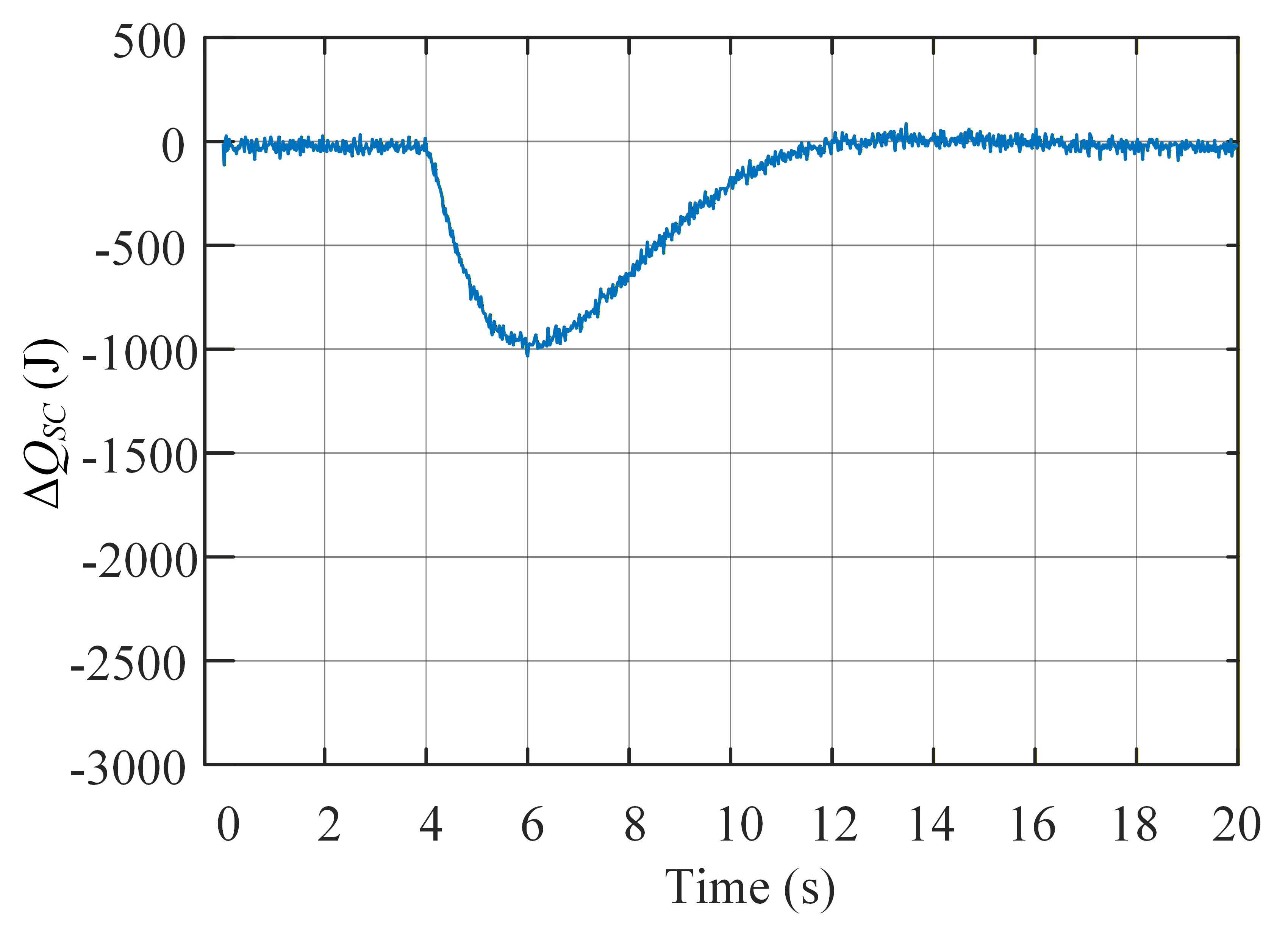}
	\caption{\textcolor{black}{Experimental results of $\Delta Q_{\mathrm{SC}}$.}}
	\label{Fig. 22}
	\vspace{-1em}
\end{figure}

\vspace{-1em}
{\color{black}
\subsection{Case 4: Hardware-Based Experimental}
In addition to the HIL tests, a prototype system was also developed to further validate the controller's performance, as shown in Fig.~\ref{Fig. 19}. The setup uses an AC programmable source to emulate the grid and a laboratory-scale DC/AC inverter board. The inverter’s PWM signals and control code are generated by a dSPACE DS1202 platform. The PEMEL, AEL, and SC branches are emulated using bidirectional programmable DC power supplies (IT6000D series). The entire test process is supervised by a host PC, and key waveforms are recorded using an oscilloscope. The system parameters are kept consistent with those used in the earlier HIL experiments.

As shown in Fig.~\ref{Fig. 20}, the grid frequency steps down from approximately 50Hz to 49.78Hz. Fig.~\ref{Fig. 21} illustrates the power responses of the three HHESS branches. Before the disturbance, both the PEMEL and AEL absorb approximately 4.65kW of power, while the SC remains at 0kW under steady-state conditions. When the frequency drops, the SC quickly discharges to supply power, the PEMEL promptly reduces its absorbed power, and the AEL responds more slowly. The system eventually reaches a new steady state, where both the PEMEL and AEL absorb around 2.8kW and the SC power returns to approximately 0kW. Fig.~\ref{Fig. 22} shows the energy variation of the SC branch. The SC energy change, $\Delta Q_{\mathrm{SC}}$, is obtained by integrating the measured SC power in MATLAB. The results indicate that $\Delta Q_{\mathrm{SC}}$ first decreases to approximately -1.0kJ during the transient support phase, and then gradually returns close to zero. This observation confirms the effectiveness of the autonomous SOC recovery mechanism within the HHESS.
}

\vspace{-0.5em}
\section{Conclusion}
{\color{black}
This paper proposes a novel HHESS and corresponding control strategy to achieve effective dynamic power allocation and enhanced frequency regulation. An inverter with inertia emulation control dynamically adjusts the total power delivered to the HHESS in response to grid frequency disturbances. AEL, PEMEL, and SC are respectively assigned with the static V-P droop, DID, and CID. Thanks to the coordination of three droop controllers, within the HHESS, high-frequency transient power is autonomously handled by SC, characterized by rapid response capability, while middle-frequency power is regulated by PEMEL, and low-frequency steady-state power is addressed by AEL, which is characterized by lower costs and longer operational lifespans. Meanwhile, SC provides immediate transient power support, significantly alleviating dynamic stresses on electrolyzers and autonomously restoring its own SOC without external communication. Implementing SOC recovery control enables the SC to withstand up to ten times more charge–discharge cycles compared to an SC without SOC recovery. A rigorous large-signal stability criterion based on MPT is also established, clearly defining system parameter boundaries to ensure robust stability. Comprehensive dynamic analyses and in-house HIL experiments verify the feasibility and effectiveness of the proposed HHESS structure and control methodology.
}

% or


% \appendices

% \section{}

%\section*{Acknowledgment}

\ifCLASSOPTIONcaptionsoff
  \newpage
\fi



%\bibliographystyle{IEEEtran}

\vspace{-0.5em}
\bibliographystyle{IEEEtran}
\bibliography{reference}

%\begin{thebibliography}{1}


%\end{thebibliography}


%\end{IEEEbiography}


%\newpage



%\vfill


\end{document}